\documentclass[useAMS, usenatbib, usegraphicx, twocolumn,letterpaper]{mn2e}

\input epsf
\usepackage{graphicx}
\usepackage{color}
\usepackage{amsfonts}
\usepackage[pdftitle={An efficient implementation of massive neutrinos in
  non-linear structure formation simulations}]{hyperref}

\newcommand{\adsurl}[1]{\href{#1}{ADS}}

\newcommand{\beq}{\begin{equation}}
\newcommand{\eeq}{\end{equation}}
\newcommand{\barr}{\begin{eqnarray}}
\newcommand{\earr}{\end{eqnarray}}

\newcommand{\rme}{\textrm{e}}

\newcommand{\bs}{\mathbf}

\newcommand{\gadget}{{\small GADGET\,}}

\newcommand{\Mpch}{\,\mathrm{Mpc} \,h^{-1}}
\newcommand{\hMpc}{h^{-1}\,\mathrm{Mpc}}
\newcommand{\Lya}{Lyman-$\alpha\;$}

\newcommand{\lesssim}{\mathrel{\hbox{\rlap{\lower.55ex\hbox{$\sim$}} \kern-.3em \raise.4ex \hbox{$<$}}}}
\newcommand{\gtrsim}{\mathrel{\hbox{\rlap{\lower.55ex\hbox{$\sim$}} \kern-.3em \raise.4ex \hbox{$>$}}}}

\title[Massive neutrinos in non-linear structure
formation simulations]{An efficient implementation of massive neutrinos in
  non-linear structure formation simulations}
\author[Y. Ali-Ha\"{\i}moud and S. Bird]{
       Yacine Ali-Ha\"{\i}moud\thanks{E-mail: yacine@ias.edu} and Simeon Bird\thanks{E-mail: spb@ias.edu}\vspace{1.5mm}\\
Institute for Advanced Study, Einstein Drive, Princeton, New Jersey 08540}

\begin{document}

\date{\today}

\pagerange{\pageref{firstpage}--\pageref{lastpage}} \pubyear{2012}
\pagenumbering{arabic}
\label{firstpage}

\maketitle

\begin{abstract}

Massive neutrinos make up a fraction of the dark matter, but due to
their large thermal velocities, cluster significantly less than 
cold dark matter (CDM) on small scales. An accurate theoretical modelling of
their effect during the non-linear regime of structure formation is
required in order to properly analyse current and upcoming
high-precision large-scale structure data, and constrain the neutrino mass. Taking advantage of the
fact that massive neutrinos remain linearly clustered up to late
times, this paper treats the linear growth of neutrino overdensities
in a non-linear CDM background. The evolution of the CDM component is
obtained via $N$-body computations. The smooth neutrino component is
evaluated from that background by solving the
Boltzmann equation linearised with respect to the neutrino
overdensity. CDM and neutrinos are simultaneously evolved in time,
consistently accounting for their mutual gravitational influence. This
method avoids the issue of shot-noise inherent to particle-based
neutrino simulations, and, in contrast with standard Fourier-space methods, 
properly accounts for the non-linear potential wells in which the 
neutrinos evolve. Inside the most massive
late-time clusters, where the escape velocity is larger than the neutrino
thermal velocity, neutrinos can clump non-linearly, causing the method
to formally break down. It is shown that this does not affect the total matter power spectrum, which can be very
accurately computed on all relevant scales up to the present time.

\end{abstract}


\section{Introduction}

The determination of neutrino masses lies on the boundary between particle physics,
astrophysics and cosmology. Atmospheric and solar neutrino oscillations have allowed the
measurement of two mass-squared differences $\Delta m_{12}^2 \equiv
m_2^2 - m_1^2 = 7.54^{+0.46}_{-0.39} \times 10^{-5}$ eV$^2$ and $|\Delta m_{13}^2| \equiv |m_3^2
- (m_1^2+m_2^2)/2| = 2.43^{+0.12}_{-0.16} \times 10^{-3}$ eV$^2$
(central values and 2-$\sigma$ error bars from \citealt{Fogli_2012}). These measurements do not allow us to
pin down the individual neutrino masses, nor their hierarchy, since
only the absolute value of $\Delta m_{13}^2$ is measured -- note,
however, that next-generation oscillation experiments such as {\textsc NOVA}\footnote{NOVA proposal: \url{http://www-nova.fnal.gov/}}
aim to distinguish the neutrino hierarchy through second-order mixing
effects. Current $\beta$-decay experiments imply an upper limit on the individual neutrino mass
$m_{\nu} \lesssim 2$ eV \citep{Kraus_2005}, and future experiments
like KATRIN should lower this limit by an order of magnitude \citep{Eitel_2005}. 

Cosmological probes are sensitive to the
absolute mass of neutrino species (see, for example, \citealt{Lesgourgues_2006} and
\citealt{Wong_review} for recent reviews), and so far provide the tightest limits
on the total neutrino mass $M_{\nu} \equiv \sum m_{\nu} \equiv m_1 +
m_2 + m_3$, at the level of a few tenths of electron-volts (eV). 
Increasingly sensitive future experiments aim to reach levels of precision 
at which a total neutrino mass $\sum m_{\nu} \sim 0.1$ eV or lower would 
be detectable (we refer the reader
to \citealt{Abazajian} for a recent compilation of current constraints
and forecasts). Ultimately, the neutrino hierarchy may even be 
within reach \citep{Takada_Komatsu_2006, Jimenez}.

Massive neutrinos affect the growth of perturbations through two effects. 
First, they change the background evolution. For a total neutrino mass less than a few eV, 
neutrinos are still relativistic at the time of matter-radiation
decoupling but non-relativistic today. At fixed total density of
non-relativistic matter today, the epoch of matter-radiation equality is
moved to later times as the neutrino mass is increased. If one holds
the density of baryons and CDM fixed instead, the angular-diameter
distance and time of matter -- dark energy equality are changed for
different neutrino masses. In each case the CMB anisotropy power
spectrum is affected \citep{Lesgourgues_2006, Wong_review}, which has allowed CMB observations to set what is perhaps the most robust
constraint on the neutrino mass, $\sum m_{\nu} < 1.3$ eV
\citep{WMAP7}. Upcoming data from the \emph{Planck} satellite is
expected to lower this bound by a factor of two \citep{Planck, Abazajian}.

The second effect of massive neutrinos is to slow down the growth of
structure on scales smaller than their free-streaming length. This
effect has long been understood in the linear regime 
\citep[and references therein]{Bond_1980, Bond_Szalay_1983, Ma_1995, Lesgourgues_2006}, 
and is implemented in all modern Boltzmann codes (see for example
\citealt{CAMB_neutrinos, CLASS_neutrinos}). For a total neutrino
mass of a few tenths of eV, this effect becomes manifest at scales
$\lesssim 100 h^{-1}$ Mpc and gets stronger in the non-linear
regime at scales of a few tenths to a few tens of Mpc
\citep{Lesgourgues_2006}. 

Combined with CMB anisotropy measurements, various probes of the
matter distribution have already constrained the total neutrino
mass to $\sum m_{\nu} \lesssim 0.2 - 0.3$ eV. These include galaxy surveys such as SDSS
\citep{SDSS_2012, Xia_2012}, cosmic shear surveys such as CFHT-LS
\citep{Ichiki_2009}, Lyman-$\alpha$ forest measurements
\citep{Seljak_2006, Viel_2010}, and galaxy cluster
surveys \citep{Vikhlinin_2009}. As the survey samples grow larger, the
sensitivity to the neutrino mass is expected to reach $\sum m_{\nu}
\sim 0.1$ eV \citep{Abazajian} or even lower (see
\citealt{Takada_Komatsu_2006} for a detailed Fisher-matrix forecast of the
constraining power of future high-redshift surveys). In addition, future measurements of
CMB lensing \citep{Hall_2012} and 21 cm surveys \citep{Mao_2008} have the potential to
reach yet lower bounds. Massive neutrinos also affect the growth of 
halos \citep{Brandbyge_2010a, Marulli_2011}.

Each one of the aforementioned methods has its own advantages and
limitations. Systematic errors may arise from using biased tracers of
the underlying density field, from difficulties in modelling complex baryonic
physics, or from foreground contamination. In addition, this level of precision 
requires an accurate modelling of corrections from non-linear structure growth.
Here we focus on the latter problem. 
Cosmological neutrinos only interact gravitationally and, 
because of their large thermal velocities, effectively 
constitute a hot dark matter (HDM) component. 
While it poses significant practical difficulties, 
the non-linear clustering of massive
collisionless particles is now a relatively well-understood 
problem. The growth of pure cold dark matter (CDM) structure can be computed 
through collisionless $N$-body simulations, whose accuracy is limited primarily
by available computational resources, which have been steadily increasing 
over time (see e.g. Fig.~1 of \citealt{Nbody}). 
By contrast, the effect of neutrinos on the non-linear growth of matter perturbations has only recently been
investigated in depth. This delicate problem has been approached
from two angles. On the one hand, extensions to perturbation theory
allow for semianalytic calculations of the CDM power spectrum in the
weakly non-linear regime. Combined with a linear treatment of
neutrinos, such methods can reach a few percent accuracy up to $k \sim
1~ h$ Mpc$^{-1}$ at high redshifts ($z \gtrsim 2$) \citep{Saito_2008,
  Saito_2009,Lesgourgues_2009, Wong_2008}. \cite{Shoji_Komatsu_2009}
go beyond linear theory for neutrinos and solve for both the CDM and
neutrino overdensities to third order in perturbation theory, making
however the simplifying assumption that neutrinos can be described by
an ideal fluid with a constant Jeans scale. \cite{Hannestad_2012} 
incorporated neutrinos into N-body simulations using a similar fluid approach.

On the other hand, the ``exact'' solution to the problem of CDM+neutrino clustering
can in principle be obtained from $N$-body simulations where both CDM
and neutrinos are simulated as particles. This approach was recently
taken by \cite{Brandbyge_2008}, \cite{Viel_2010} and \cite{Bird_2012} (see
also references therein for earlier works). The main limiting factor
of such an approach is numerical: since neutrinos do not significantly
cluster below their free-streaming scale, their power spectrum on small 
scales is dominated by shot noise for any reasonable number of simulated 
particles\footnote{The shot noise issue arises for neutrinos for two reasons. First, their
  large random velocities effectively randomly distribute them over
  the box, which leads to a Poisson spectrum $P(k) = 1/\overline{n}$
  at all scales. Second, their intrinsic clustering is very small on
  small scales. At large enough redshifts and for small enough scales,
the true power may be completely swamped by the Poisson noise.},
increasingly so for lower neutrino masses and at earlier times. A vast
amount of computational power and memory is therefore wasted to extract 
a small amount of information on the actual neutrino clustering. 
In order to bypass the shot-noise issue, \cite{Brandbyge_2009}
included neutrino perturbations in Fourier space, while still
treating the CDM component as particles in $N$-body
simulations. The density field of their neutrino component was only computed using linear
theory, however, and did not account consistently for non-linear CDM
perturbations sourcing the gravitational potential in which neutrinos
evolve \citep{Bird_2012}. Recently, \cite{Brandbyge_2010} suggested a
hybrid method, where neutrinos are initially treated using the Fourier method with
linear theory, and converted to particles as time progresses. One of
the remaining sources of errors of this implementation is that the
Fourier-space part does not account for the non-linear growth of gravitational
potentials in the neutrino evolution.

The aim of the present work is to close this gap and provide an efficient method for accurately
computing the effect of massive neutrinos on the non-linear matter power spectrum,
with a minimal modification to well-tested pure CDM $N$-body simulation methods. 
To do so, we follow the CDM with the Tree-PM $N$-body code
\textsc{Gadget-3} \citep{Springel_2005, Viel_2010}, and
compute analytically the linearised neutrino overdensity sourced by the full gravitational potential. The CDM and neutrinos are evolved
simultaneously and self-consistently, accounting for their mutual
gravitational influence. Our method is therefore semi-linear, in the
sense that we account exactly for the non-linear growth of CDM overdensities
and gravitational potentials, but effectively use a
linear transfer function to obtain the neutrino overdensity from the
gravitational potential. For neutrinos of the mass allowed by current data, 
our implementation is sufficient on its own for a complete
description of the matter power spectrum up to the present time, and, for $z \geq 1$, the 
neutrino power spectrum as well. It only fails to resolve neutrino clustering in the most massive galaxy clusters, in which
neutrinos do cluster non-linearly \citep{Ringwald_Wong_2004}; however, it should be easy to 
perform zoomed simulations of these clusters using our method as a base. An ancillary advantage of our implementation is that it can
easily be added into an $N$-body code as a small ``patch'' and does not require running an independent Boltzmann code in parallel.
Furthermore, our code can easily include the exact contribution of
neutrinos to the background expansion, and the effect of the neutrino hierarchy, which are problematic or expensive to 
include in particle implementations. It can be applied to regimes where shot noise 
severely limits the reliability of the particle method, such as low neutrino masses, the $21$cm forest, or 
large-volume galaxy mock catalogue simulations. Perhaps the most important property of our code 
is that it allows massive neutrinos to be included into simulations with 
negligible cost in CPU and memory, and with no loss of accuracy. Thus it can be useful for interpreting 
the results of essentially any probe of large scale structure, including CMB lensing, the 
Lyman-$\alpha$ forest, weak lensing experiments or 
galaxy surveys \citep{Takada_Komatsu_2006, Cooray_1999, Kaplinghat_2003,Wang_2005, Vallinotto_2009, Gratton_2007,Ichiki_2009,Schlegel_2009, Carbone_2011,Euclid_2010}.

This paper is organised as follows. In Section \ref{sec:notation}, we
define our notation and discuss characteristic scales and the regime
of validity of our main approximation. We derive our main equations in Section
\ref{sec:Equations}. The practical interfacing with our $N$-body code
is detailed in Section \ref{sec:implementation}. We describe our
results and compare them against other methods in Section
\ref{sec:results}. We discuss potential applications of our method in
Section \ref{sec:applications} and conclude in Section \ref{sec:conclusion}.

\section{Notation, Characteristic timescales and
  lengthscales}\label{sec:notation}

\subsection{Notation}
Throughout this paper we use $\tau$ for the conformal time ($d\tau
= dt/a$, where $t$ is the physical time and $a = (1+z)^{-1}$ is the scale factor) and work in units where $c = G
= k_{\rm B} = 1$. Lengthscales and wavenumbers are comoving unless otherwise
stated. Overdots denote differentiation with respect to conformal time
and gradients are with respect to comoving lengths. The Hubble rate
today is $H_0 = 100~ h$ km/s/Mpc. When referring to an individual neutrino mass, we
use the lower-case $m_{\nu}$. When referring to the sum of neutrino
masses, we use the upper-case $M_{\nu} \equiv \sum m_{\nu}$. The
temperature of the (unperturbed) neutrino background today is 
\beq
T_{\nu,0} = 1.95 \textrm{ K} = 1.68 \times 10^{-4} \textrm{ eV}.
\eeq
Neutrinos being non-relativistic today, they contribute a fraction
$f_{\nu}$ of the total nonrelativistic matter abundance, given by
\beq
f_{\nu} = \frac{\Omega_{\nu}}{\Omega_{\rm M}} = \frac1{\Omega_{\rm M} h^2} \frac{\sum
  m_{\nu}}{93.14 ~\rm{eV} } \approx 0.022 ~\frac{0.147}{\Omega_{\rm M}
  h^2}~ \frac{\sum m_{\nu}}{0.3 \rm eV}.
\eeq

\subsection{Relativistic to nonrelativistic transition}
We denote by $\bs p$ the physical momentum of a neutrino and $f_0(\bs
p)$ the unperturbed neutrino phase-space density, giving the number of neutrinos and antineutrinos per unit
physical phase-space in the absence of gravitational clustering. Neutrinos decouple from the baryon plasma at a temperature of about 1
MeV \citep{Lesgourgues_2006}. Given that their masses are known to be
less than 1 eV, this means that they were ultrarelativistic at
decoupling. This implies that their unperturbed phase-space density is
the redshifted relativistic Fermi-Dirac distribution (since $\bs p \propto a^{-1}$
in the absence of gravitational potentials),
\beq
f_0(\bs p) = \frac2{h^3} \frac{1}{\exp(p/T_{\nu}) + 1}, \label{eq:Fermi-Dirac}
\eeq 
with $T_{\nu}(z) = (1+z)T_{\nu,0}$. A neutrino of mass $m_{\nu}$
becomes non-relativistic when its momentum falls below its mass, $p \lesssim
m_{\nu}$. This occurs at a redshift $z_{\rm nr}$ such that
\beq
1 + z_{\rm nr} \approx \frac{m_{\nu}}{T_{\nu,0}} \frac{T_{\nu}}p \approx
595 \frac{m_{\nu}}{0.1 \textrm{ eV}} \frac{T_{\nu}}p.
\eeq
The Fermi-Dirac distribution (\ref{eq:Fermi-Dirac}) is such that 50\%
of neutrinos have a momentum $p < 2.84 ~ T_{\nu}$ and 90\%
have a momentum $p < 5.47 ~ T_{\nu}$. As a consequence, 50\% of
neutrinos are nonrelativistic for $z \lesssim 200~(m_{\nu}/0.1
\textrm{ eV})$ and 90\% have become non-relativistic by $z \lesssim 100~(m_{\nu}/0.1
\textrm{ eV})$. For the lowest masses
considered ($m_{\nu} \approx 0.05$ eV), relativistic corrections
to neutrino clustering may be of order unity at high redshifts (we
start our simulations at $z = 49$); however, neutrinos are basically
unclustered on all scales when they are quasi-relativistic, and the
exact value of their very small inhomogeneities is then irrelevant for CDM
clustering. We therefore treat neutrinos in the non-relativistic
limit, i.e. assume $p \ll m_{\nu}$ in all our
derivations\footnote{Note that consistently accounting for
relativistic corrections in the evolution of neutrinos would also
require accounting for CMB inhomogeneities sourcing the gravitational
potentials; these are always (rightfully so) neglected in $N$-body simulations.}. 

\subsection{Free-streaming scale}

\begin{figure}
\includegraphics[width=0.45\textwidth]{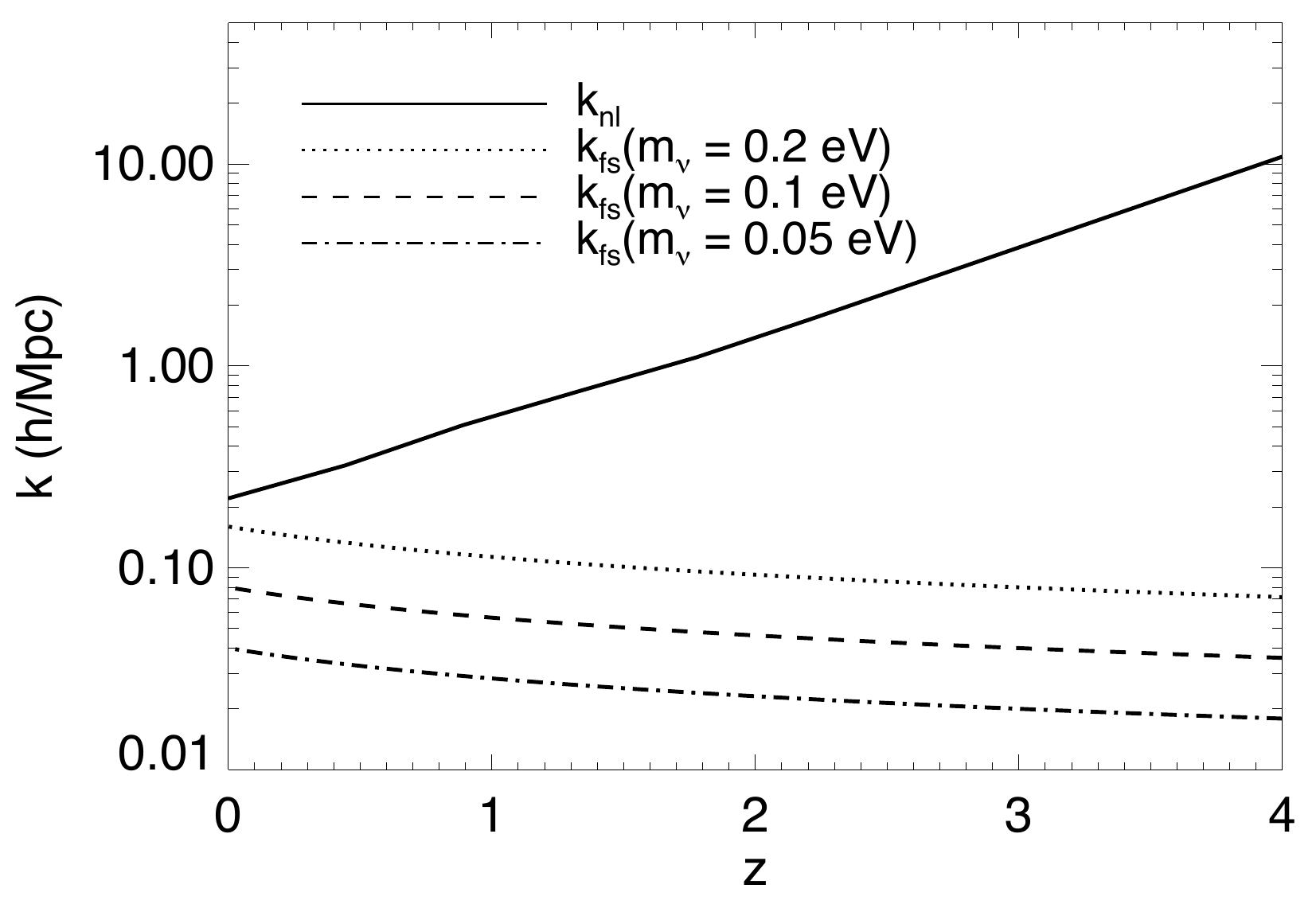}
\caption{Non-linear scale (solid) and free-streaming scales for various
  neutrino masses, as a function of redshift.}
\label{fig:kfs-knl}
\end{figure}

Neutrinos can free-stream across a
comoving lengthscale $\lambda$ if the time it takes them to cross $\lambda$ is
much less than the Hubble time $H^{-1}$, i.e. if
\beq
\frac{a \lambda}{v_{\nu}} \ll H^{-1},
\eeq
where $v_{\nu}(z) \sim T_{\nu}(z)/m_{\nu}$ is the characteristic neutrino velocity.
This equation defines a characteristic comoving \emph{free-streaming scale}
$k_{\rm fs}(z) \sim a H/v_{\nu}(z)$. A more detailed analysis in Section \ref{sec:Equations} will
show that it is convenient to define the free-streaming scale as 
\beq
k_{\rm fs}(a) \equiv \left(\frac32 \Omega_{\rm M}(a)
  \overline{v^{-2}}\right)^{1/2} a H, \label{eq:kfs-formal}
\eeq
where $\overline{v^{-2}}$ is the mean inverse velocity squared,
\beq
\overline{v^{-2}} = \frac{2 \ln(2)}{3 \zeta(3)}
\left(\frac{m_{\nu}a}{T_{\nu, 0}} \right)^2 
\approx \left( 810~ \textrm{km/s} ~(1+z)\frac{0.1~ \rm eV}{m_{\nu}}\right)^{-2}, \label{eq:v2}
\eeq
 and $\Omega_{\rm M}(a)$ is the relative contribution of the
non-relativistic components to the total energy density at scale
factor $a$. On scales much larger than the free-streaming scale,
neutrinos behave like CDM. We shall show later on that for $k \gg k_{\rm
  fs}$, the (linearised) neutrino overdensity $\delta_{\nu}$ relates
to the total (possibly non-linear) matter overdensity
$\delta_{\rm M}$ as 
\beq
\delta_{\nu}(k \gg k_{\rm fs}) \approx \left(\frac{k_{\rm
      fs}}{k}\right)^2 \delta_{\rm M}.
\eeq 
Numerically, we obtain
\beq
 k_{\rm fs}(z) \approx \frac{0.08}{\sqrt{1+z}}
\sqrt{\frac{\Omega_{\rm M}}{0.3}} \frac{m_{\nu}}{0.1 ~ \textrm{eV}} h~ \textrm{Mpc}^{-1}.  \label{eq:kfs}
\eeq
We show the non-linear scale $k_{\rm nl}$ (defined such that the
variance of CDM overdensity per log-$k$ interval reaches unity on that
scale) and free-streaming scale as a
function of neutrino mass and redshift in Fig.~\ref{fig:kfs-knl}. We
see that for $\sum m_{\nu} < 0.6$ eV, we always have $k_{\rm fs} <
k_{\rm nl}$, with an increasing difference at large redshifts. Moreover, the non-linear matter power spectrum typically
grows as $k^3 P_{\rm M}(k)\propto k^{\alpha}$, with $\alpha \lesssim 2$
\citep{Seljak_2000} and therefore we expect the power per log interval in the
neutrino component to be decreasing for $k \gtrsim k_{\rm nl} > k_{\rm
fs}$ as $k^3 P_{\nu}(k) \propto k^{\beta}$, with $\beta \lesssim -2$.

We therefore expect the neutrino component to be linearly clustered on all scales
for masses below the current upper limit: for $k < k_{\rm nl}$
neutrinos cluster at most like the CDM, which is itself linearly clumped. For $k >
k_{\rm nl} > k_{\rm fs}$, the neutrino power per logarithmic interval is a decreasing function
of $k$. 

This argument motivates our use of
linear theory for the neutrino component. However, this does not give
the full physical picture of neutrino clustering: the slowest
neutrinos may in fact significantly cluster in massive haloes. Before
moving to the core of our calculation, we first discuss in which
conditions it may break down.

\subsection{Neutrino capture in massive haloes} \label{sec:clusters}
 In this section we qualitatively discuss under which conditions
 neutrinos may significantly cluster in a potential well of
 characteristic \emph{physical} scale $r_0$ and characteristic depth
 $\phi_0 < 0$. Let us consider, for simplicity, a spherical top hat potential, $\phi(r < r_0) = \phi_0$ and $\phi(r > r_0) = 0$. We assume that $r_0$ is constant in time, but allow the
 depth $\phi_0$ to vary slowly, on a Hubble timescale, as is the case
 for the most massive halos currently forming.  
We consider the fate of a particle on a purely radial trajectory. If the particle enters the
potential at time $t_i$, with initial velocity
$v(t_i^-) = v_i$, its velocity upon entry becomes
\beq
v(t_i^+) = \sqrt{v_i^2 + 2 |\phi_0(t_i)|}.
\eeq
Since we have assumed the potential to be flat inside the halo, the
particle's velocity is conserved until it reaches the other end, at
time $t_f$. By then the gravitational potential has grown a little
deeper, and the particle will escape only if its velocity is larger
than the new escape velocity, i.e. if 
\beq
v_i^2 + 2 |\phi_0(t_i)| > 2 |\phi_0(t_f)|.
\eeq
Provided the crossing time is short compared to the
evolution timescale of the potential, we can Taylor-expand this
equation and obtain the escape condition
\beq
v_i^2 > 2 (t_f - t_i) |\dot{\phi}_0|.
\eeq
Inside the halo, provided $2|\phi_0| \gg v_i^2$, the velocity is
approximately $\sqrt{2 |\phi_0|}$ and the crossing time is therefore
\beq
t_f - t_i \approx \frac{2 r_0}{\sqrt{2 |\phi_0|}}.
\eeq
If we define the timescale for variation of $\phi$ as 
\beq
\Delta t_{\phi} \equiv \frac {|\phi_0|}{|\dot{\phi}_0|},
\eeq
the escape condition for neutrinos becomes
\beq
r_0 < \frac12 (H \Delta t_{\phi}) \frac{v}{H} \frac{v}{\sqrt{2|\phi_0|}}, \label{eq:escape-cond}
\eeq
where we have purposefully inserted the Hubble parameter to make the order-unity
parameter $H \Delta t_{\phi}$ appear.
We see that for deep potential wells varying on the Hubble timescale, the condition for neutrinos to
truly free-stream and not be captured is more stringent than simply
requiring their characteristic scale to be smaller than the (physical) free-streaming scale $r_{\rm
  fs} \sim v/H$. Equation (\ref{eq:escape-cond}) can also be turned
into an escape condition for the neutrino momentum. For $z \approx 0$,
this is
\barr
\frac{p}{T_{\nu}} \gtrsim  \frac{m_{\nu}}{T_{\nu,
    0}} \frac{1}{\sqrt{H_0 \Delta t_{\phi}}} \left( 2 H_0 ~r_0 \sqrt{|2 \phi_0|}\right)^{1/2}~~~~~~~~~~~~~~~~~~~~~~~~\nonumber\\
\approx \left(H_0 \Delta t_{\phi}\right)^{-1/2} \frac{m_{\nu}}{0.1~ \rm eV} \left(\frac{r_0}{0.5 ~ h^{-1} \rm
  Mpc}\right)^{1/2} \left(\frac{\sqrt{|\phi_0|}}{3000 ~ \rm{km/s}}\right)^{1/2}.\label{eq:escape-analytic}
\earr
We have normalised the lengthscale and depth of the gravitational well
to values typical for the most massive haloes. The outcome of this analysis
is that in massive haloes varying on a Hubble timescale, neutrinos with momentum $p \lesssim
T_{\nu}$ are typically captured, while those with momentum $p \gtrsim
T_{\nu}$ can escape. We emphasise that the cutoff value at $p \approx T_{\nu}$ is a
pure numerical coincidence, arising from the characteristic sizes and
depths of massive haloes,
and for neutrino masses of order 0.1 eV. 
Given that only about 6\% of neutrinos have a momentum $p < T_{\nu}$, the qualitative picture that emerges
from this analysis is that a relatively small fraction of neutrinos are efficiently
bound to massive haloes, thereafter strongly clustering, while a
majority remain weakly clustered. 

Let us emphasise that although the crossing time of a neutrino in a massive halo is significantly
shorter than the Hubble time (for the fiducial values of
Eq.~(\ref{eq:escape-analytic}), $H \delta t \approx 1/40$ at $ z= 0$),
the relevant comparison is not of $H \delta t$ with unity but
rather with the small ratio $v^2/\phi$ (also $\approx 1/40$ for our adopted
fiducial values): even a small relative effect due to the change
of the potential on a Hubble timescale may translate into an order unity
effect on the pre-entry velocity and prevent the neutrino from
escaping the halo.

The analysis presented here is of course very simplified, for we have used an
idealised (and unphysical) profile for the gravitational potential. 
Moreover, in reality there is no sharp boundary in momentum space between linear
behaviour and strong clustering. It is
however a robust statement to say that the characteristic momentum of
neutrinos efficiently captured in massive haloes is of order of their
temperature. If the critical momentum were much larger than
$T_{\nu}$, the bulk of neutrinos would be captured and the neutrino overdensity in massive clusters would
reach values comparable to those for the CDM. If it were much smaller
than $T_{\nu}$, then linear theory would be extremely accurate for
neutrinos even in the vicinity of massive clusters. The actual
situation is intermediate between these two regimes, as shown in the
work of \cite{Ringwald_Wong_2004}, who used $N$-one-body simulations
to track the evolution of neutrinos around massive clusters. They found that for $m_{\nu} =
0.15$ eV, linear theory underestimates neutrino clustering by a factor
of $\sim 2$ near the centre of $10^{14} M_{\odot}$ clusters, and by a factor of
$\sim 4$ for $10^{15} M_{\odot}$ clusters; for $M \leq 10^{12}
M_{\odot}$, they found an excellent agreement with linear theory for
$m_{\nu} = 0.15$ eV.

With this caveat in mind, we now proceed to the main thrust of this
paper: linear perturbations of neutrinos in a general gravitational potential.

\section{Linear evolution of hot dark matter perturbations in a general background} \label{sec:Equations}

In this section we consider the linear evolution of an ensemble of
non-relativistic hot dark matter (HDM) particles of mass $m$ in a general gravitational 
potential $\phi$ in an expanding universe. We derive an integral equation known as the Gilbert equation
\citep{Gilbert_1966}, which has been used in various analytic works on
neutrinos \citep{Bond_Szalay_1983, Brandenberger_1987, Singh_Ma_2003,
  Abazajian_Switzer_2005}. This equation can easily be generalised to the
relativistic case \citep{Weinberg_book, Shoji_Komatsu_2010, Vega1, Vega2}, although it
takes on a more complicated form.

\subsection{Vlasov equation}

\subsubsection{Derivation for non-relativistic particles in an expanding universe}

In principle, the Boltzmann equation in an expanding universe should be
derived in a consistent relativistic fashion (see for example
\citealt{Ma_1995}). However, for non-relativistic particles and in the
Newtonian ($\phi \ll 1$) and subhorizon ($k \gg a H$) limits, which always hold on the scales of
interest, we can derive it more simply as follows.

We choose an arbitrary coordinate origin $\bs{x}_0$ and denote by $\bs r =
a(\tau)(\bs x - \bs{x}_0)$ the physical coordinate of
a particle, where $\bs x$ is its comoving coordinate. We also denote
by $\bs p$ the physical momentum of the particle,
\beq
\bs p \equiv m \frac{d \bs r}{d t} = m H \bs r + m \frac{d \bs x}{ d \tau}
\equiv m H \bs r + a^{-1} \bs q,
\eeq
where the last equality defines the comoving momentum $\bs q \equiv a
m ~d \bs x/d \tau$. The equation of motion for a test particle is 
\beq
\frac{d \bs p}{d t} = - m \bs{\nabla}_{\bs r} \Phi, \ \ \textrm{i.e.} \ \
\frac{d \bs q}{d \tau} = - m a \bs{\nabla}_{\bs x} \Phi - m a
\frac{d^2 a}{d t^2} \bs r, \label{eq:dot.q}
\eeq
where $\Phi$ is the total gravitational potential. We now decompose
$\Phi$ into a piece due to the background density and a piece sourced
by the density perturbations:
\beq
\Phi = \Phi_0 + \phi,
\eeq
where 
\barr
\Phi_0(\bs r) &=& -\frac1{2a}\frac{d^2 a}{d t^2} \bs{r}^2 = -\frac12
\left(H^2 + \frac{d H}{dt}\right) \bs{r}^2\nonumber\\
&=& \frac{2\pi}{3}\left(\overline{\rho} + 3 \overline{p}\right) \bs{r}^2
\earr
and $\phi$ is determined from the Poisson equation
\beq
\nabla_{\bs r}^2 \phi = 4 \pi \delta \rho.
\eeq
Equation (\ref{eq:dot.q}) for the comoving momentum then simplifies to
\beq
\frac{d \bs q}{d \tau} = - m a \bs{\nabla}_{\bs x} \phi. \label{eq:dot.q.2}
\eeq
We now define $f(\bs x, \bs q, \tau)$ as the HDM phase-space density, 
which is the number of particles per unit of physical
phase-space (i.e. $dN_{\rm hdm} = f(\bs x, \bs q, \tau) d^3 r d^3p$). In the absence of collisions, the conservation of phase-space along a
particle's trajectory gives the collisionless Boltzmann equation (also
known as the Vlasov equation):
\beq
\left.\frac{d f}{d \tau}\right|_{\rm traj} = \frac{\partial f}{\partial \tau} +
\frac{d \bs x}{d \tau} \cdot\bs{\nabla}_{\bs x} f + \frac{d \bs q}{d
  \tau}\cdot\bs{\nabla}_{\bs q} f = 0. 
\eeq
Finally, using the definition of $\bs q$ and Eq.~(\ref{eq:dot.q.2}) for
its derivative, we find (independently of the chosen origin $\bs{x}_0$): 
\beq
\frac{\partial f}{\partial \tau} + \frac{\bs q}{m a} \cdot\bs{\nabla}_{\bs
  x} f - m a  \bs{\nabla}_{\bs x} \phi \cdot \bs{\nabla}_{\bs q} f = 0.
\eeq

\subsubsection{Linearization and Fourier transform}

In a homogeneous universe, $\bs{\nabla}_{\bs x} \phi = 0$ and
$\nabla _{\bs x}f =
0$. As a consequence, the solution to the Vlasov equation must satisfy
$\partial f_0/\partial \tau = 0$, i.e. be a function of momentum
only. Finally, isotropy guarantees that $f_0 = f_0(q)$ is a function
of $q \equiv |\bs q|$ only. 

We now linearise the Vlasov equation about the homogeneous solution:
$f(\bs x, \bs q, \tau) = f_0(q) + \delta f(\bs x, \bs q, \tau)$,
assuming the perturbation $\delta f \ll f_0$ is induced by the presence
of the gravitational potential $\phi$ (our formal expansion parameter
is $(m a)^2 \phi/q^2$). The linearised equation becomes
\citep{Brandenberger_1987, Ma_1995}
\beq
\frac{\partial \delta f}{\partial \tau} + \frac{\bs q}{m a} \cdot\bs{\nabla}_{\bs
  x} \delta f - \frac{m a}{q} \frac{d
  f_0}{d q}  \bs{q} \cdot \bs{\nabla}_{\bs x} \phi  = 0.\label{eq:linear-real}
\eeq
We can now Fourier-transform this equation and obtain
\beq
\frac{\partial \delta f}{\partial \tau} + i \frac{\bs{q} \cdot \bs
  {k}}{ma} \delta f = i \frac{m a}{q} \frac{d
  f_0}{d q} (\bs{q} \cdot \bs k) \phi, \label{eq:Fourier}
\eeq
where $\bs k$ is the comoving wavenumber, and $\delta f(\bs k, \bs q, \tau)$ and $\phi(\bs k, \tau)$ should
now be understood as the Fourier transforms of $\delta f$ and $\phi$, respectively.

\subsection{Integral solution}

For given comoving wavenumber and momentum, Eq.~(\ref{eq:Fourier}) is
a linear first order ordinary differential equation, and has the
explicit solution
\barr
\delta f(\bs k, \bs q, \tau) = \rme^{- i \frac{\bs q \cdot \bs k}{m}(s
- s_i)} \delta f(\bs k, \bs q, \tau_i)~~~~~~~~~~~~~~~~~~~~~~~~\nonumber\\
 +~ i\frac{m}{q} \frac{d
  f_0}{d q} (\bs{q} \cdot \bs k) \int_{\tau_i}^{\tau} \rme^{- i \frac{\bs{q} \cdot \bs{k}}{m}(s
- s')} a(\tau') \phi(\bs k, \tau') d \tau', \label{eq:delta.formal}
\earr
where $\tau_i$ is some initial conformal time, and we have used the variable (sometimes referred to as the
``superconformal time'')
\beq
s(\tau) \equiv \int_{\tau_i}^{\tau} \frac{d \tau'}{a(\tau')}.
\eeq
We can now evaluate the average of $\delta f$ over the
direction of momentum $\hat{q}$ (this will be needed shortly):
\barr
&&\overline{\delta f}(\bs k, q, \tau) \equiv \frac{1}{4 \pi} \int d^2
\hat{q} \delta f(\bs k, \bs q, \tau)\\
&&= \overline{\delta f}^I(\bs k, q, \tau_i, \tau)\nonumber\\
&& + ~m \frac{d f_0}{d q} k \int_{\tau_i}^{\tau} j_1\left(k\frac{q}{m}(s
- s')\right) a(\tau') \phi(\bs k, \tau') d \tau', \label{eq:deltaf.av}
\earr
where $j_1$ is the order-one spherical Bessel function of the first
kind (see for example \citealt{Bessel}), and we have defined
\beq
\overline{\delta f}^I(\bs k, q, \tau_i, \tau) \equiv \frac1{4 \pi} \int d^2 \hat{q} ~\rme^{- i \frac{\bs q \cdot \bs k}{m}(s
- s_i)} \delta f(\bs k, \bs q, \tau_i). \label{eq:deltafI.av}
\eeq
The mass density of the considered particle can be evaluated
by integrating the phase-space density over momenta:
\beq
\rho_{\rm hdm}(\bs x, \tau) = m a^{-3} \int d^3 q  f(\bs x, \bs q,
\tau). \label{eq:rho}
\eeq
We can now obtain the
density perturbation in Fourier space,
\barr
\delta \rho_{\rm hdm}(\bs k, \tau) &=& m a^{-3}\int d^3 q \delta f(\bs k,
\bs q, \tau)\nonumber\\
 &=&  m a^{-3}\int 4 \pi q^2 dq ~\overline{\delta
  f}(\bs k, q, \tau).
\earr
Dividing by the mean density $\overline{\rho}_{\rm hdm}$ (obtained from
Eq.~(\ref{eq:rho}) with $f = f_0$), using Eq.~(\ref{eq:deltaf.av}) and integrating the second term by
parts, we obtain the following expression for the density contrast $\delta_{\rm hdm} \equiv
\frac{\delta \rho_{\rm hdm}}{\overline{\rho}_{\rm hdm}}$:
\barr
&&\delta_{\rm hdm}(\bs k, \tau) = \delta^I_{\rm hdm}(\bs k, \tau_i, \tau)\nonumber\\
&&~- k^2 \int_{\tau_i}^{\tau}  (s - s') ~\mathcal{I}\left[\frac{k}{m}(s -
  s')\right] a(\tau')\phi(\bs k, \tau') d \tau'. ~~~~~~\label{eq:delta.almost}
\earr  
In the above equation, $\delta_{\rm hdm}^I(\bs k, \tau_i, \tau)$ is the value of the density
contrast evolved from $\tau_i$ to $\tau$ in the absence of
any gravitational potential,
\beq
\delta^I_{\rm hdm}(\bs k, \tau_i, \tau) \equiv \frac{\int dq q^2
  \overline{\delta f}^I(\bs k, q, \tau_i, \tau) }{\int
  dq ~q^2 f_0(q)}. \label{eq:deltaI}
\eeq
If we expand the phase-space density
at the initial time on the basis of Legendre polynomials (see for
example \citealt{Legendre}), 
\beq
\delta f(\bs k, \bs q, \tau_i) = \sum_{l=0}^{\infty}i^l \delta f_l(\bs k, q, \tau_i) P_l(\hat{k} \cdot \hat{q}),
\eeq
we obtain the following expression for $\delta^I$: 
\beq
\delta^I_{\rm hdm}(\bs k, \tau_i, \tau) = \sum_{l=0}^{\infty}\frac{\int dq~q^2 \delta f_l(\bs k, q, \tau_i)j_l\left(\frac{k
q}{m}(s - s_i)\right)}{\int dq q^2 f_0(q)}. \label{eq:deltaI-multipole}
\eeq
The dimensionless function $\mathcal{I}$ in Eq.~(\ref{eq:delta.almost}) is the Fourier transform of the
unperturbed distribution function in momentum space, normalised so
that $\mathcal{I}(0) = 1$ \citep{Brandenberger_1987, Bertschinger_Watts_1988},
\beq
\mathcal{I}[X; f_0] \equiv \frac{\int dq~ j_0(q X) q^2 f_0(q) }{\int
  dq ~q^2 f_0(q)}. \label{eq:I.def}
\eeq
For neutrinos described by a relativistic Fermi-Dirac distribution, we
provide an accurate fitting formula for this function in Appendix \ref{sec:fitIX}.

Finally, since the gravitational potential is sourced by the total matter density 
through the Poisson equation
\beq
k^2 \phi = - 4 \pi a^2 \overline{\rho}_{\rm M} \delta_{\rm M} = - \frac32 H_0^2 \frac{\Omega_{\rm M}}{a} \delta_{\rm M},\label{eq:Poisson}
\eeq
where $\Omega_{\rm M}$ is the matter fraction at the present day, we can rewrite Eq.~(\ref{eq:delta.almost}) as
\barr
&&\delta_{\rm hdm}(\bs k, \tau) = \delta^I_{\rm hdm}(\bs k, \tau_i, \tau)\nonumber\\
&&~+ \frac32  H_0^2 \Omega_{\rm M} \int_{\tau_i}^{\tau} \mathcal{I}\left[\frac{k}{m}(s -
  s')\right] (s - s') \delta_{\rm M}(\bs k, \tau') d \tau'. ~~~~~~\label{eq:delta.final}
\earr  
Note that $\delta_{\rm hdm}$ at time $\tau$ depends, in principle, on
its value at all prior times through $\delta_{\rm M} (\tau' <
\tau)$.

\subsection{Discussion: limiting regimes}
The unperturbed phase-space density is in general characterised by a
typical comoving momentum $q_0$, such that $f_0(q)$ decreases rapidly
for $q \gg q_0$. For neutrinos, for example, $q_0 = T_{\nu,0}$. We also assume that this is the case for $\delta
f$; indeed, provided this is true initially, it will remain true as long as linear theory is valid,
as can be seen from Eq.~(\ref{eq:delta.formal}).  

\subsubsection{Large scales}\label{sec:large-scales}

In the limit $k q_0(s - s_i)/m \ll 1 $, we may Taylor-expand the
spherical Bessel functions and obtain, up to terms of $\mathcal{O}(k^2)$, 
\barr
\delta_{\rm hdm}(k \rightarrow 0, \tau) &=& \delta_{\rm
  hdm}(\bs k, \tau_i) - a_i \theta_{\rm hdm}(\bs k,
  \tau_i) (s - s_i)\nonumber\\
 &-& k^2 \int_{\tau_i}^{\tau} a(\tau') (s - s') \phi(\bs k, \tau') d
 \tau', \label{eq:fluid.sol}
\earr
where $\delta_{\rm hdm}(\bs k, \tau_i)$ is the initial overdensity:
\beq
\delta_{\rm hdm}(\bs k, \tau_i) \equiv \frac{\int d q q^2 \delta
  f_0(\bs k, q, \tau_i)}{\int d q q^2 f_0(q)}\,,
\eeq
and $\theta_{\rm hdm}(\bs k, \tau_i)$ is the initial bulk velocity
divergence:
\beq
\theta_{\rm hdm}(\bs k, \tau_i) \equiv \frac13 k \frac{\int d q q^2
  (q/m a_i) \delta
  f_1(\bs k, q, \tau_i)}{\int d q q^2 f_0(q)}\,.
\eeq
Equation (\ref{eq:fluid.sol}) is nothing but the explicit solution to
the (Newtonian, sub-horizon) linearised fluid equations in an expanding universe (see for
example \citealt{Ma_1995}):
\barr
\dot{\delta}_{\rm hdm} + \theta_{\rm hdm} &=& 0, \\
\dot{\theta}_{\rm hdm} + \frac{\dot{a}}{a} \theta_{\rm hdm} &=& k^2 \phi.
\earr
We therefore recover the well-known fact that on scales much larger
than their free-streaming length, HDM particles behave as
a pressureless ideal fluid. 

In general, the HDM component should behave like a cold species on
scales much larger than its free-streaming length, regardless of whether it is linear. 
It is therefore important, for our linear approximation to
work across all scales, that the CDM is indeed linear on scales larger than the free-streaming
length, otherwise we would not get the correct long-wavelength
behaviour. This is ensured by the hierarchy of scales $k_{\rm fs} <
k_{\rm nl}$.

\subsubsection{Small scales}\label{sec:small-scales}

Let us define the function 
\beq
E(\tau, \bs x, \bs q) \equiv \frac{q^2}{2 m} + m a^2 \phi(\tau, \bs x).
\eeq
The change of $E$ along a particle's trajectory is
\beq
\frac{d E}{d \tau}\big{|}_{\rm traj} = \frac{\partial E}{\partial
  \tau} + \frac{d \bs x}{d \tau} \cdot \frac{\partial E}{\partial
  \bs x} + \frac{d \bs q}{d \tau} \cdot \frac{\partial E}{\partial
  \bs q} = \frac{\partial E}{\partial \tau},
\eeq
where the partial time derivative is at constant $\bs x$ and $\bs q$
and we have used the geodesic equations for $\bs x, \bs q$ to cancel
out the last two terms. We therefore see that 
\beq
\left|\frac{d E}{d \tau}\big{|}_{\rm traj} \right| \lesssim
a H E,
\eeq
provided the potential varies on a Hubble timescale. This is an upper
limit: if $m a^2|\phi| \ll q^2/m$ then the rate of change of $E$ can
in fact be much smaller than $ a H E$.

Let us now consider a perturbation with scale small enough that the crossing time is much
shorter than the Hubble time (which is itself of order of or shorter
than the timescale over which $E$ changes). In that case, the quantity
$E$ is an adiabatic invariant during the crossing of the
perturbation. We assume that any particle within
the perturbation can be traced back to a large distance from the
perturbation at some earlier time $\tau_i$, where $m a_i^2
\phi(\tau_i, \bs x_i) \ll q_i^2/m$ and hence $E_i = E \approx q_i^2/(2
m)$ (the distance must however be small enough to be crossed in a
timescale short compared to the time for $E$ to change significantly). If furthermore at that initial time the phase-space density is
nearly unperturbed, $f(\tau_i) \approx f_0(q_i)$, we deduce, using
conservation of phase-space, that, for $E > 0$, 
\barr
f(\tau, \bs x, \bs q) &\approx& f_0(\bs q_i) \approx f_0(\sqrt{2 m E})
\nonumber\\
 &=& f_0\left(\sqrt{q^2 + 2 (m a)^2 \phi(\tau, \bs x)}\right), \label{eq:f-adiabatic}
\earr
and for $E < 0$, $f(\tau, \bs x, \bs q) = 0$. This argument is generic to small-scale perturbations and does not
assume linearity\footnote{Interestingly, for a relativistic
  Fermi-Dirac distribution, the small-scale overdensity
  obtained from the linearised version of Eq.~(\ref{eq:f-adiabatic}) is
  \emph{larger} than the one obtained from the full non-linear equation.}.

If we now assume that $|\phi| \ll q^2/(m a)^2$, we can linearise
Eq.~(\ref{eq:f-adiabatic}) in $\phi$ and obtain
\beq
\delta f(\tau, \bs x, \bs q) \approx \frac{(m a)^2}{q} \frac{d
  f_0}{d q} \phi(\tau, \bs x).
\eeq
We find the resulting overdensity, after integrating $\delta f$ by parts over momenta, 
\beq
\delta_{\rm hdm} \approx - \overline{v^{-2}} \phi, \label{eq:small-scales}
\eeq
where $\overline{v^{-2}}$ is the mean inverse velocity squared,
\beq
\overline{v^{-2}} \equiv (m a)^2 \frac{\int dq f_0(q)}{\int dq q^2 f_0(q)}.
\eeq
Note, in passing, that these expressions require that $df_0/dq
\rightarrow 0$ as $q\rightarrow 0$, and would not be valid for a
Bose-Einstein distribution, for example. 

Therefore, in the regime of shallow potentials, clustering on small scales is proportional to the square of
the ratio of the characteristic velocity dispersion of the potential to a
characteristic velocity of the HDM particles. Using the Poisson
equation (\ref{eq:Poisson}), we may also rewrite this in Fourier space
as
\beq
\delta_{\rm hdm}(k) \approx \frac{3}{2} k^{-2} a^2 H^2 \overline{v^{-2}} \Omega_{\rm
  M}(a) \delta_{\rm M} \equiv \left( \frac{k}{k_{\rm
      fs}}\right)^{-2} \delta_{\rm M}(k),
\eeq
where we have used the definition of the free-streaming scale,
Eq.~(\ref{eq:kfs-formal}) and $\Omega_{\rm M}(a)$ is again the relative contribution of the
non-relativistic components to the total energy density at scale
factor $a$. This simple dependence is what motivated the exact numerical prefactors used in the
definition of $k_{\rm fs}$. This result was previously derived by
\cite{Ringwald_Wong_2004}. A similar result was obtained for baryon
clustering on scales much smaller than the Jeans scale by
\cite{Gnedin_1998} (in that case the Jeans scale plays the role of
free-streaming scale, even though physically the absence of clustering is due to
the pressure response rather than free-streaming as is the case for neutrinos).

We could also have used our integral expression
Eq.~(\ref{eq:delta.final}) to reach the same
results: if $k \gg k_{\rm fs}$, then $(i)$ $k q/m(s - s_i) \gg 1$ and
the initial conditions become irrelevant, $\delta^{\rm I} \rightarrow
0$, and $(ii)$ the function $\mathcal{I}$ decreases rapidly, such
that only the times $\tau' \approx \tau$ matter in the integral. In
that case one finds, after a change of variables,
\beq
\delta_{\rm hdm}(k \gg k_{\rm fs}) \approx - (m a)^2 \phi
\int_0^{\infty} X \mathcal{I}(X) d X, 
\eeq
which can be shown to give precisely Eq.~(\ref{eq:small-scales}).

\section{Application to simulating massive neutrinos}
\label{sec:implementation}

Equation (\ref{eq:delta.final}) allows us to compute the density field 
for hot dark matter from a possibly non-linear dark matter potential. 
We shall now specialise to the case where the hot dark matter is a
neutrino component interacting gravitationally with cold dark matter and baryons, 
whose non-linear evolution is computed using an $N$-body code.

\subsection{Initial conditions}

The initial unperturbed distribution function for massive neutrinos which decoupled 
while relativistic is the Fermi-Dirac distribution,
\beq
f_0(q) = \frac{g}{h^3}\frac{1}{\rme^{q/T_{\nu,0}} + 1},
\eeq
where $g = 2$, the degeneracy of the neutrino species. Only the first
two moments of the perturbed distribution function at the initial time
are relevant: on small scales the initial conditions are rapidly
forgotten, and on large scales the higher-order moments scale as $v^l$
and decay rapidly with $l$ as neutrinos are non-relativistic. In
principle, one should extract the full momentum information from a
Boltzmann code at the startup redshift, and obtain $\delta f_0(\bs k,
q, \tau_i)$ and $\delta f_1(\bs k, q, \tau_i)$. However, on large
scales, neutrinos essentially behave as a cold species, with local
overdensities and bulk flows roughly independent of their individual
momenta, and on small scales, the exact form of initial conditions is
irrelevant. These considerations allow us to assume the following
simple form for the initial distribution function:
\barr
f_{\nu}(\bs x, \bs q, \tau_i) = f_0(|\bs q-m_{\nu} a_i\bs v_{\nu}(\bs x, \tau_i)|)[1 +
\delta_{\nu}(\bs x, \tau_i)]~~~~~\nonumber\\
 ~~~~~~~\approx f_0(q)\left[1 + \delta_{\nu}(\bs
  x, \tau_i) - a_i m_{\nu} \bs v_{\nu}(\bs x, \tau_i) \cdot \hat q \frac{d \ln f_0}{d q}\right],
\earr
where we have assumed that the bulk velocity $\bs{v}_{\nu}$ is smaller than the
characteristic random velocity. Fourier-transforming this equation,
and using $\bs{v}_{\nu}(\bs k) = - \frac{i}{k} \hat{k} \theta_{\nu}(\bs k)$,
we obtain the multipole moments of our approximate initial phase-space
density:
\barr
\delta f_0(\bs k, q, \tau_i) &=& f_0(q) \delta_{\nu}(\bs k, \tau_i),\\
\delta f_1(\bs k, q, \tau_i) &=& \frac{d f_0}{d q} m_{\nu}
k^{-1} a_i\theta_{\nu}(\bs k, \tau_i),\\
\delta f_l(\bs k, q, \tau_i) &=& 0, \ \  \ l \geq 2.
\earr
The initial conditions propagate to the current redshift
according to Eq.~(\ref{eq:deltaI-multipole}), which gives us,
after integrating the second term by parts,
\beq
\delta^I(\bs k, \tau_i, \tau) = \mathcal{I}_{s_i, s} \left[\delta_{\nu}(\bs k, \tau_i) - a_i
  \theta_{\nu}(\bs k, \tau_i) (s - s_i) \right],
\label{eq:deltaI-nu}
\eeq
where $\mathcal{I}_{s_1, s_2} \equiv \mathcal{I}([s_2 -
s_1]k/m )$ and $\mathcal{I}$ was defined in Eq.~(\ref{eq:I.def}).

\subsection{Phase evolution}

Storing the full three-dimensional Fourier transform of the
gravitational potential at each time step in
order to perform the integral in Eq.~(\ref{eq:delta.final}) would
require prohibitive amounts of memory. We therefore make some simplifying 
assumptions so that we need only store the power spectrum.

Previous works \citep{Viel_2010, Brandbyge_2009} used the fully-linear
neutrino overdensity (i.e. computed assuming even the CDM is linear) and assumed that 
the random phases and relative amplitudes of the neutrino density field did not evolve, 
but were given by the initial conditions. This is true on linear scales, where 
there is no mode mixing, but not on non-linear scales, where neutrinos,
even if linear themselves, evolve in the gravitational potential
sourced by non-linear CDM perturbations, for which mode-mixing is important.

Here again, we can take advantage of the hierarchy of free-streaming and non-linear scales $k_{\rm fs}(z) <
k_{\rm nl}(z)$ for currently allowed neutrino masses. For linear scales $k < k_{\rm nl}(z)$, the CDM phases are
approximately constant and equal to the phase at the current time
$\tau$ for any $\tau' < \tau$. For scales smaller than the
free-streaming scale, the integral in Eq.~(\ref{eq:delta.final}) is 
dominated by the most recent times, such that $|\tau' - \tau| \ll \tau$. It
is reasonable to assume that phase evolution and mixing are not too
important during a short period of time, and therefore the phases at
all relevant times are approximately equal to the phase at the present
time. Because of the hierarchy of scales $k_{\rm fs}(z) <
k_{\rm nl}(z)$, we can therefore assume, for all scales, that the phase and relative amplitude at $\tau' < \tau$ is equal to that at the current time $\tau$, and therefore
\beq
\delta_{\rm M}(\bs k, \tau') \approx \left(\frac{P_{\rm M}(k, \tau')}{P_{\rm M}(k, \tau)}\right)^{1/2} \delta_{\rm
 M}(\bs k, \tau), \label{eq:phases.assumption}
\eeq
where $P_{\rm M}$ is the matter power spectrum. We also assume that the initial condition piece
satisfies a similar relation 
\beq
\delta_{\nu}(\bs k, \tau_i) \approx \left(\frac{P_{\nu}(k,
    \tau_i)}{P_{\nu}(k, \tau)}\right)^{1/2} \delta_{\nu}(\bs k, \tau), \label{eq:phases.initial}
\eeq
and that the ratio $\theta_{\nu}(\bs k, \tau_i)/\delta_{\nu}(\bs k,
\tau_i) \equiv [\theta_{\nu}/\delta_{\nu}]_i(k)$ is a function of $k$
only. The former is justified because initial conditions are only
relevant on large (and linear) scales, where
Eq.~(\ref{eq:phases.initial}) is indeed correct, and the latter because
structure is fully linear at our starting redshift.

Substituting Eqs.~(\ref{eq:phases.assumption}) and Eq.~(\ref{eq:phases.initial}) into 
Eq.~(\ref{eq:delta.final}), with the initial condition piece given by
Eq.~(\ref{eq:deltaI-nu}), we obtain, first, that the neutrino component is in
phase with the CDM, and second, that the neutrino power-spectrum is
given by
\barr
P_{\nu}^{1/2}(k, \tau) &=& \mathcal{I}_{s_i, s}
P_{\nu}^{1/2}(k, \tau_i) \left\{1 - (s - s_i)  a_i [\theta_{\nu}/\delta_{\nu}]_{i}(k)\right\}\nonumber\\
&+& \frac32 \Omega_{\rm M} H_0^2 \int_{\tau_i}^{\tau} \mathcal{I}_{s', s}
P^{1/2}_{\rm M}(k, \tau') (s - s')d \tau', \label{eq:P-final}
\earr
where $\mathcal{I}_{s_1, s_2} \equiv \mathcal{I}([s_2 -s_1]k/m
)$. Because neutrino and CDM overdensities are in phase (following
from our assumptions), we can then recover the full neutrino density field from $P_{\nu}^{1/2}(k, \tau)$ 
and the CDM density field by
\beq
\delta_{\nu}(\bs k, \tau) = \left(\frac{P_{\nu}(k,
    \tau)}{P_{\rm cdm}(k, \tau)}\right)^{1/2} \delta_{\rm cdm}(\bs k, \tau).\label{eq:phases}
\eeq

\subsection{Implementation}

We implement the effect of massive neutrinos in Fourier space 
as a modification to the TreePM-SPH code \gadget-3
\citep{Springel_2005, Viel_2010}. In order to lower the
computational cost, and following \cite{Viel_2010}, we do not compute
the short-range tree force due to and experienced by neutrinos. The
size of a PM grid cell for our fiducial simulation is $0.5 ~h^{-1}$Mpc,
corresponding to a wavenumber $k \approx 13~h$
Mpc$^{-1}$, a scale two orders of magnitude smaller than the neutrino
free-streaming length. Neutrino overdensities are therefore
completely negligible compared to CDM overdensities below the grid
scale, and we are justified in neglecting the Tree force due to
them. However, neglecting the Tree force experienced by neutrinos is not
strictly justifiable, and effectively amounts to having a lower
resolution for the neutrinos than for the CDM \citep{Viel_2010}. As we shall discuss below, we checked that our results are converged 
with respect to grid size, proving that there is a negligible contribution from 
scales smaller than our grid resolution, where the tree force would be
active. 

We evolve CDM particles (and baryons) as well as neutrino overdensities simultaneously as
follows: 

$(i)$ We generate adiabatic initial conditions using power spectra from
\textsc{camb} and setting equal random relative amplitude and phases
for the CDM, baryon and neutrino initial overdensities.

$(ii)$ Given the total matter overdensity field up
to time $\tau - \Delta \tau$, we evolve the CDM and baryons with our
Tree-PM code up to the next timestep $\tau$. We then evaluate the
Fourier-transformed CDM+baryon density, as is required for computing the
long-range particle forces, and compute the CDM+baryon power
spectrum at time $\tau$. 

$(iii)$ In order to evaluate the neutrino overdensity at time $\tau$ with
Eq.~(\ref{eq:P-final}), we require the total matter power-spectrum up
to (and including) the current time $\tau$. Equation (\ref{eq:P-final}) is therefore strictly speaking an implicit equation
for $P_{\nu}(k, \tau)$, and should be solved iteratively. Since
the integrand vanishes at $\tau' = \tau$ (because of the $(s - s')$
term), and neutrinos contribute a small fraction of the dark matter,
with a good initial guess a single iteration is sufficient. We
show this in detail in Appendix
 \ref{app:iteration}. We therefore approximate $P_{\nu}(k, \tau) \approx
 P_{\nu}(k, \tau - \Delta \tau)$ to evaluate $P_{\nu}(k, \tau')$
 inside the integral of Eq.~(\ref{eq:P-final}) -- we do so by
 interpolating over our tables of stored power spectra. We checked
 that $P_{\nu}(k, \tau)$ obtained from this first iteration is
 converged at the level of $\sim 10^{-4}$ by comparing with the output
 of a second iteration, which is therefore not required.

$(iv)$ We then recover the full phase information for the neutrino
overdensity from Eq.~(\ref{eq:phases}), and evaluate the total matter
overdensity at the current timestep $\tau$ from
\beq
\delta_{\rm M}(\bs k, \tau) = (1-f_\nu) \delta_{\rm cdm, b}(\bs k, \tau) + f_\nu \delta_{\nu}(\bs k, \tau).
\eeq
We finally compute and store the total matter power spectrum at time
$\tau$, and reiterate steps $(ii)$ to $(iv)$ until the end of the simulation.

\subsection{Simulation Details} 

The parameters of our simulations are shown in Table \ref{tab:partsimuls}. 
Our cosmological parameters were similar to those in \cite{Bird_2012}. 
We set $h = 0.7$, the total matter fraction $\Omega_{\rm M} = 0.3$, 
the scalar spectral index $n_s = 1$ and the amplitude of the primordial power spectrum, 
$A_s= 2.43 \times 10^{-9}$. Our initial conditions were generated, from transfer functions 
generated by \textsc{camb}, using our own version of N-GenICs modified to 
use 2LPT \citep{Scoccimarro_1998} and account correctly for baryons in the initial transfer 
function\footnote{Our initial conditions generator is freely available at \url{http://github.com/sbird/S-GenIC}.}.
Where it was important, we used $\Omega_b = 0.05$. 

In a previous work \citep{Bird_2012} we found that a significant
limitation 
to the inclusion of massive neutrinos was the restriction on initial redshift due to the slightly relativistic behaviour 
of the neutrinos at early times. Although our method for computing the growth of neutrino 
perturbations is non-relativistic, we 
can at least easily include the correct contribution of neutrinos to the
background evolution by computing
\beq
\overline{\rho}_\nu(a) = a^{-3}\int \epsilon(q, a) f_0(q)  4 \pi q^2 
dq\,,
 \label{eq:relativistic}
\eeq
where $\epsilon(q, a) = \sqrt{m^2_\nu + a^{-2} q^2}$ is the total
energy of a neutrino with comoving momentum $q$. Thus,
$\overline{\rho}_\nu(a)$ interpolates between the behaviours of matter and radiation.
This is extremely easy to implement with our Fourier method, but
somewhat harder in a purely particle simulation, where the energy
density is by default computed from the particles' rest-mass and does not account
for their kinetic energy.
For consistency, we also included the effect of radiation density in the background evolution, and, for CDM only simulations, 
the effect of massless neutrinos. Our particle-based simulations still
have $\overline{\rho}_{\nu} = \overline{\rho}_{\nu, 0} a^{-3}$.

\begin{table}
\begin{center}
\begin{tabular}{|c|c|c|c|c|c|}
\hline
Name   & $m_\nu$ (eV) & Box ($\Mpch$) & $N_\mathrm{CDM}^{1/3}$ & $z_i$ & $N_\mathrm{bar}^{1/3}$\\
\hline
S00   & $0$    & $256$ & $512$ &  $49$  & \\
S00P   & $0$    & $256$ & $1024$ &  $49$  & \\  
S05*   & $0.05$ & $256$ & $512$ &  $49$  & \\
S10*   & $0.1$  & $256$ & $512$ &  $49$  & \\
S15*   & $0.15$  & $256$ & $512$ &  $49$ & \\
S20*   & $0.2$  & $256$ & $512$ &  $49$ & \\
S40*   & $0.4$  & $256$ & $512$ &  $49$ & \\
F10BA* & $0.1$  & $60$ & $512$ &  $49$ & $512$\\
S05IC   & $0.05$  & $256$ & $512$ &  $99$  & \\
S10P*   & $0.1$  & $256$ & $1024$ &  $49$  & \\
S03NH   & $0.03$, NH & $256$ & $512$ &  $49$  & \\
S03IH   & $0.03$, IH & $256$ & $512$ &  $49$  & \\
\hline
\end{tabular}
\end{center} 
\caption{Summary of simulation parameters. $m_\nu$ is the mass of one neutrino species.
Rows marked with a * were run using both Fourier and particle neutrinos.
Cosmological parameters were the same for all simulations and are given in the text. 
Where neutrino particles were included, the same particle number was used as for the dark matter particles. 
Simulations with $m_\nu = 0$ included massless neutrinos. 
Note we held $\Omega_{\rm M}$ constant, making $\Omega_\mathrm{cdm}$ dependent on $\Omega_\nu$.
S03NH and S03IH had the same total neutrino mass as S05, but had three species with masses 
following the normal and inverted hierarchies, respectively.
}
\label{tab:partsimuls}
\end{table}

\section{Code Validation} \label{sec:results}

\subsection{Convergence tests}
\label{sec:convergence}

\begin{figure}
\includegraphics[width=0.45\textwidth]{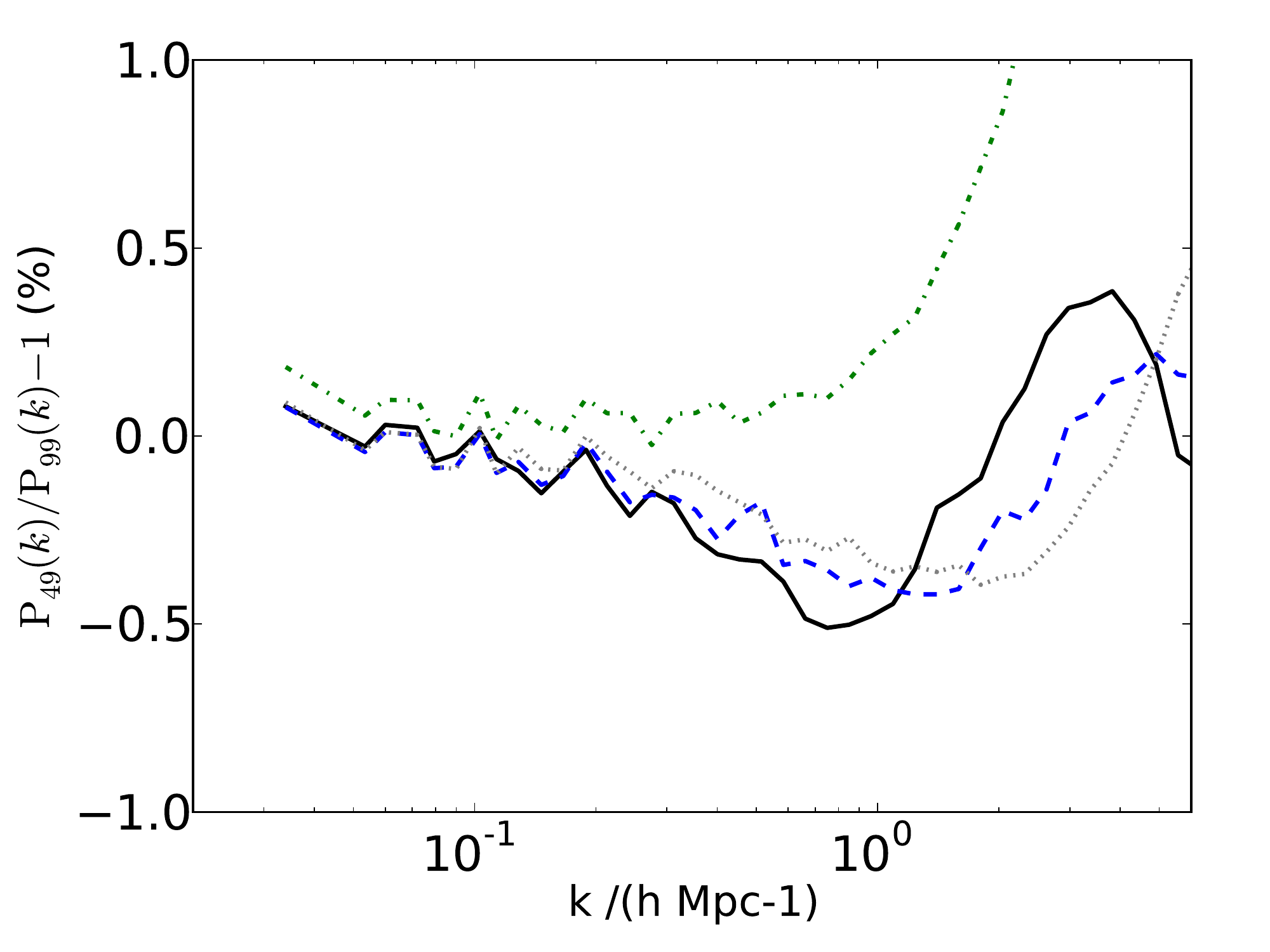}
\caption{Change in the total matter power spectrum between S05
  (initial redshift $z_i = 49$) and S05IC (initial redshift $z_i = 99$). 
Positive values correspond to more power in S05.
Each line shows a different redshift: $z=9$ (green dot-dashed), $z=3$ (grey dotted), $z=1$ (blue dashed) 
and $z=0$ (black solid). 
}
\label{fig:P_tot_IC}
\end{figure}

\begin{figure}
\includegraphics[width=0.45\textwidth]{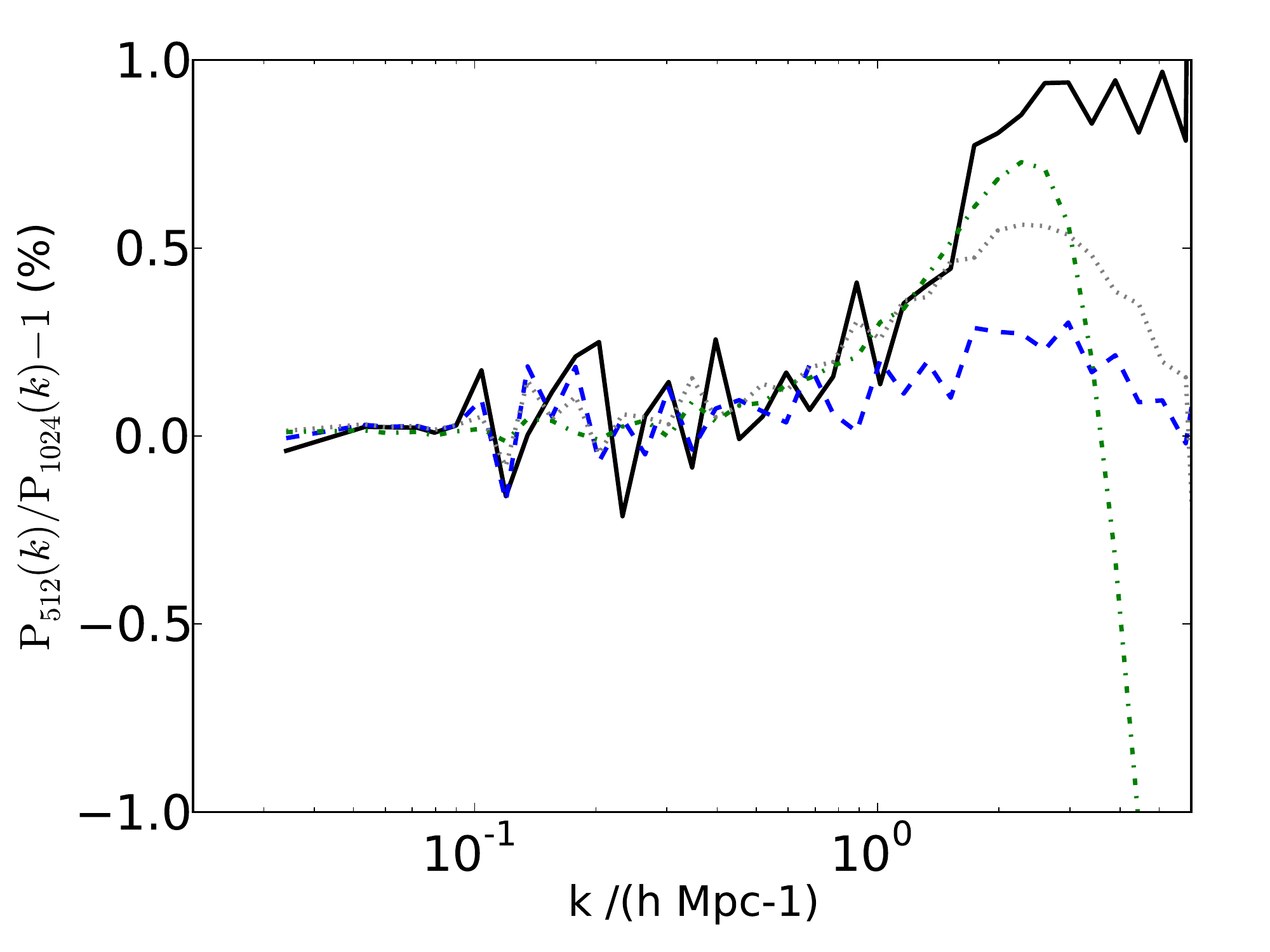}
\caption{Effect of the number of CDM particles on the power
  suppression by neutrinos. What is plotted is the fractional
  difference between the \emph{ratio}
  $(P_\mathrm{S10}/P_\mathrm{S00})$ (both using $512^3$ CDM particles) and
  the \emph{ratio} $(P_\mathrm{S10P}/P_\mathrm{S00P})$ (both using
  $1024^3$ CDM particles). Positive values correspond to a larger
  suppression of power in the $1024^3$-particle simulation.
Each line shows a different redshift: $z=9$ (green dot-dashed), $z=3$ (grey dotted), $z=1$ (blue dashed) 
and $z=0$ (black solid). 
}
\label{fig:P_tot_part}
\end{figure}

In this Section, we check the convergence of our method with respect
to particle number and initial redshift. 

Figure \ref{fig:P_tot_IC} shows the impact of changing the initial redshift from $z_i=49$ to $z_i=99$ for 
neutrinos of mass $m_\nu = 0.05$ eV. 
The change at low redshift is of order $0.5 \%$ and is consistent with slightly more small-scale non-linear growth for the 
simulation with the higher initial redshift, as expected from a pure cold dark matter simulation.

Figure \ref{fig:P_tot_part} shows the effect of changing the particle
number, comparing S10 to S10P, which has 8 times more particles. 
Changing the number of particles causes the sampling of initial structure to change slightly, introducing 
sample variance. To minimise this effect we show the ratio of the suppression due to neutrinos for pairs of simulations 
with different particle numbers, that is,
$(P_\mathrm{S10}/P_\mathrm{S00}) / (P_\mathrm{S10P}/P_\mathrm{S00P}) -
1$.
Our results are converged at the $1\%$ level for $k < 10 \hMpc$, and at the $0.2\%$ level for $k < 1 \hMpc$.
The effect of increasing particle resolution is to increase the
overall growth of non-linear structure, and as a consequence to enhance the
relative suppression due to massive neutrinos.

\subsection{Total Power Spectra}

\begin{figure}
\includegraphics[width=0.45\textwidth]{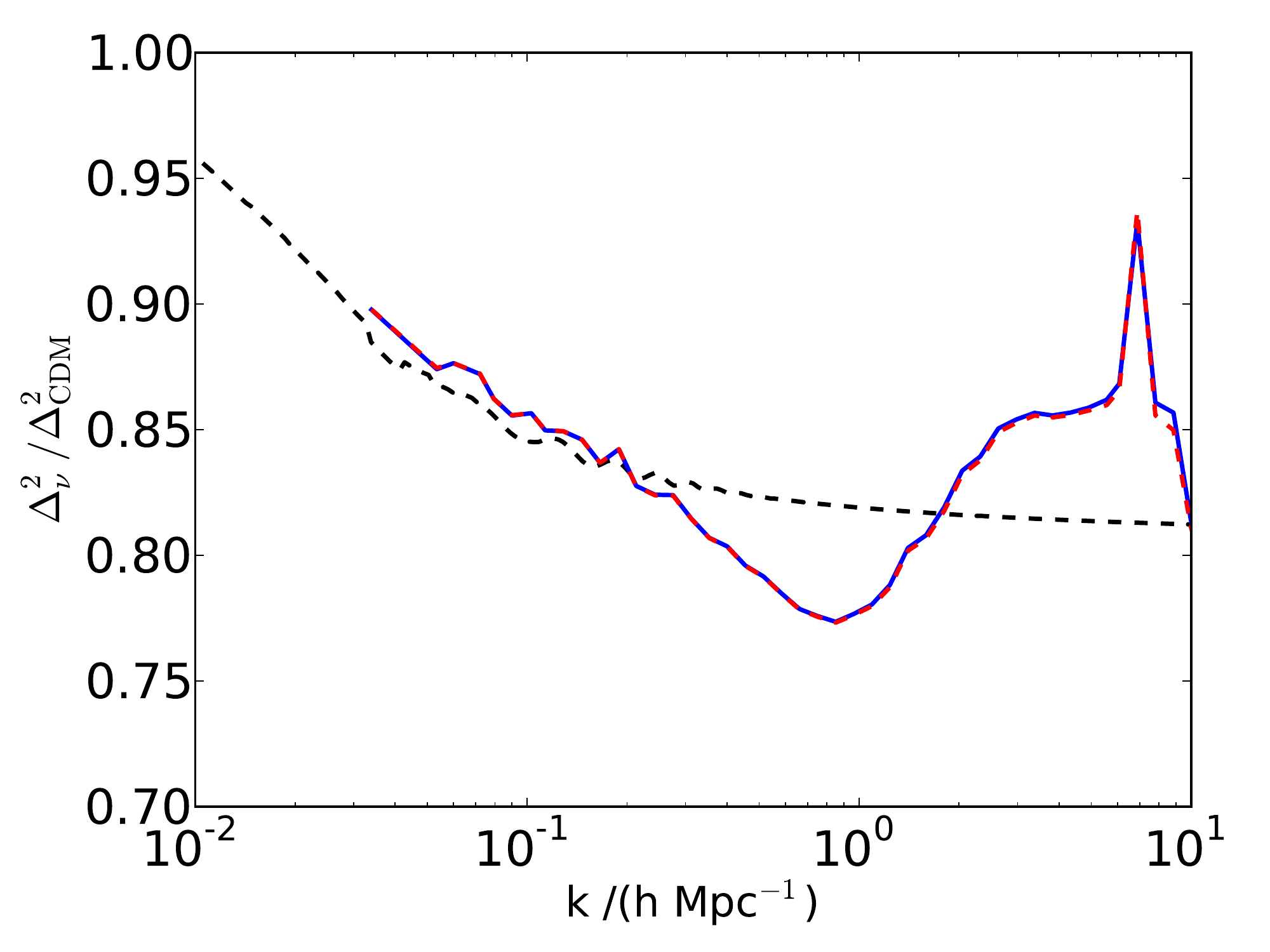}
\caption{The suppression in the total power spectrum at $z=0$.
Solid lines show a $256\Mpch$ with our fourier-based method (blue) and 
the particle method (red). The dashed line shows the prediction of
linear theory. The spike at $k \approx 7~ h/$Mpc (corresponding to
half the grid frequency) is a numerical
artefact.}
\label{fig:P_tot_z0}
\end{figure}

\begin{figure*}
\includegraphics[width=0.45\textwidth]{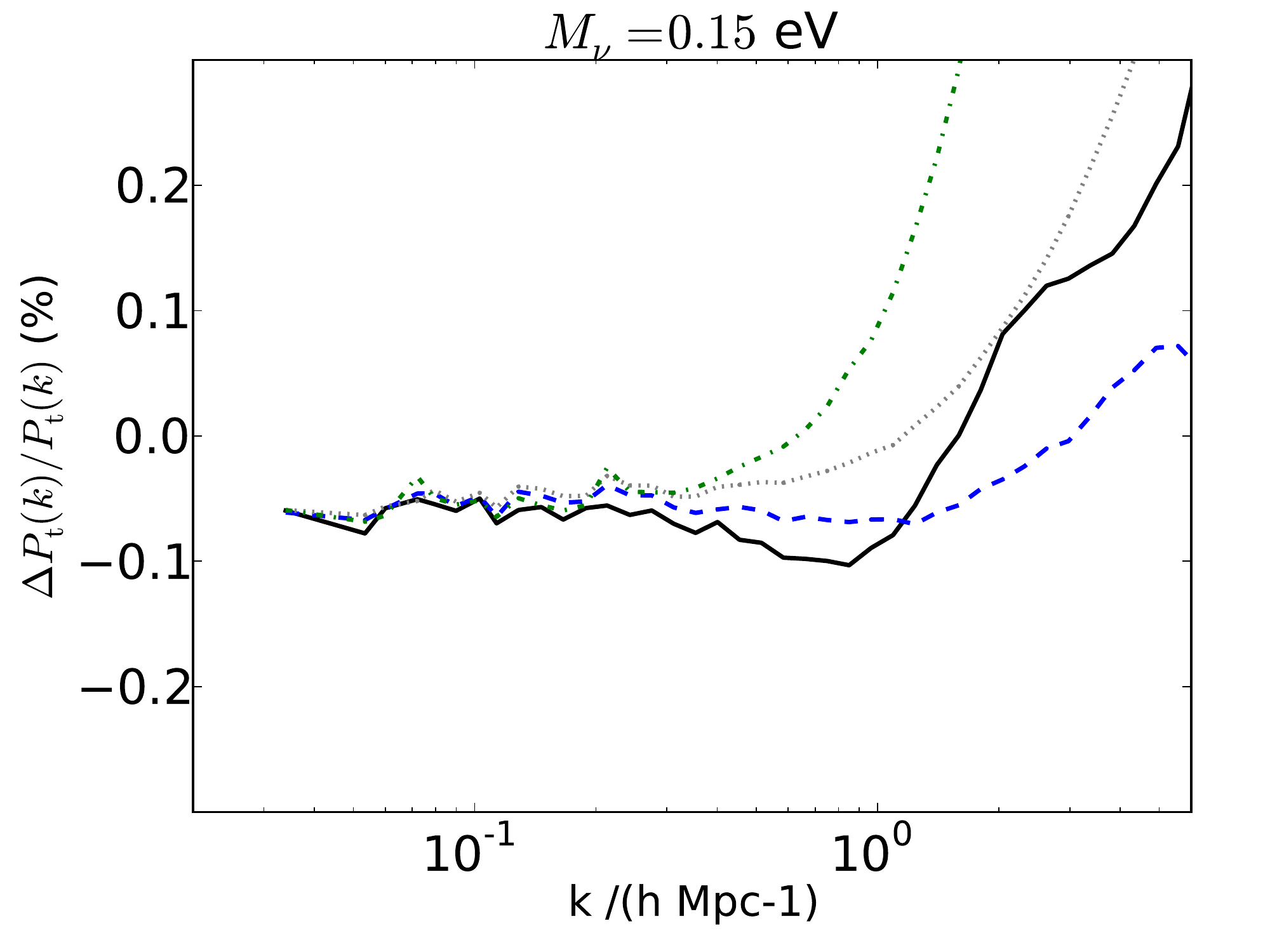}
\includegraphics[width=0.45\textwidth]{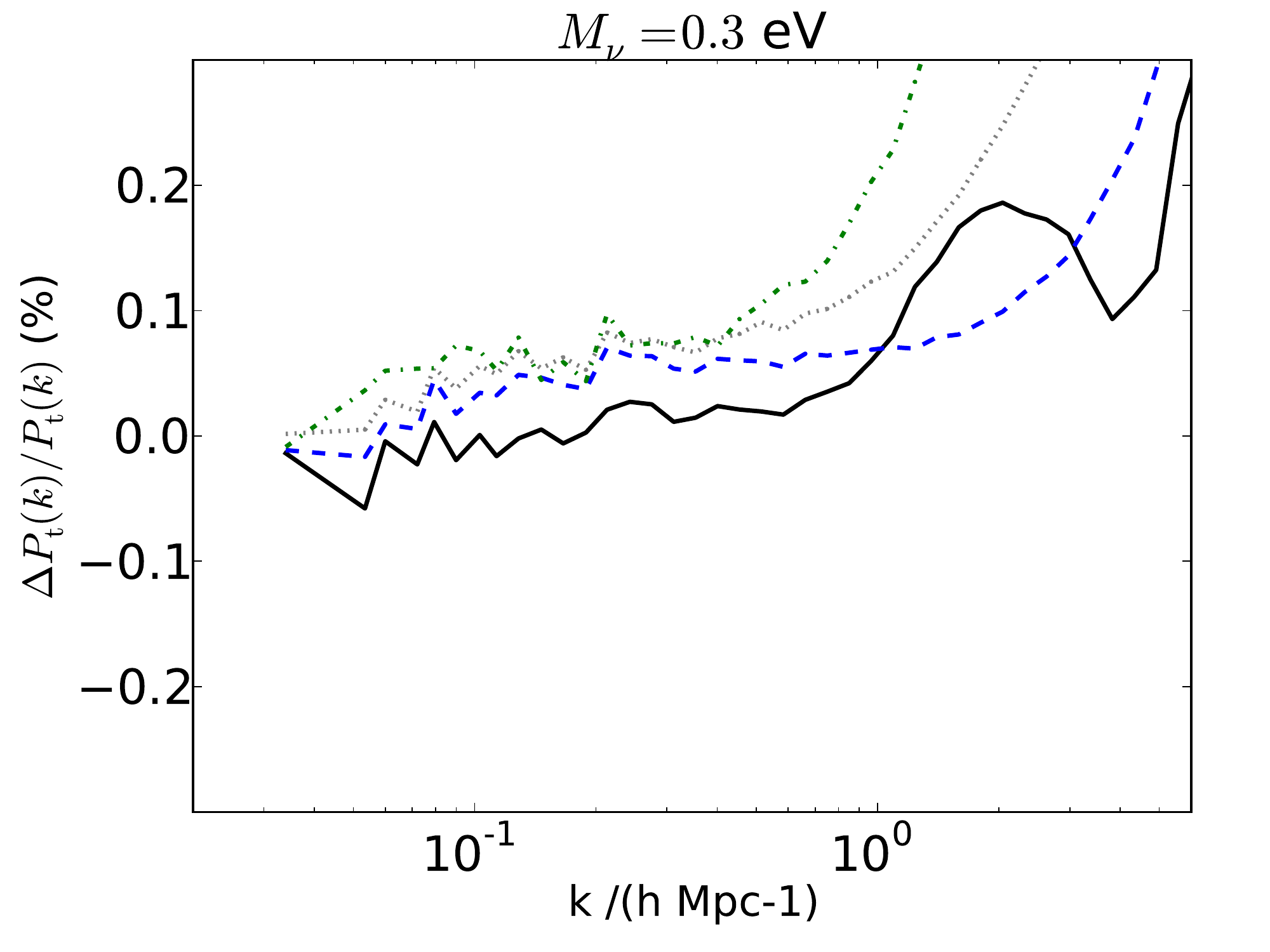} \\
\includegraphics[width=0.45\textwidth]{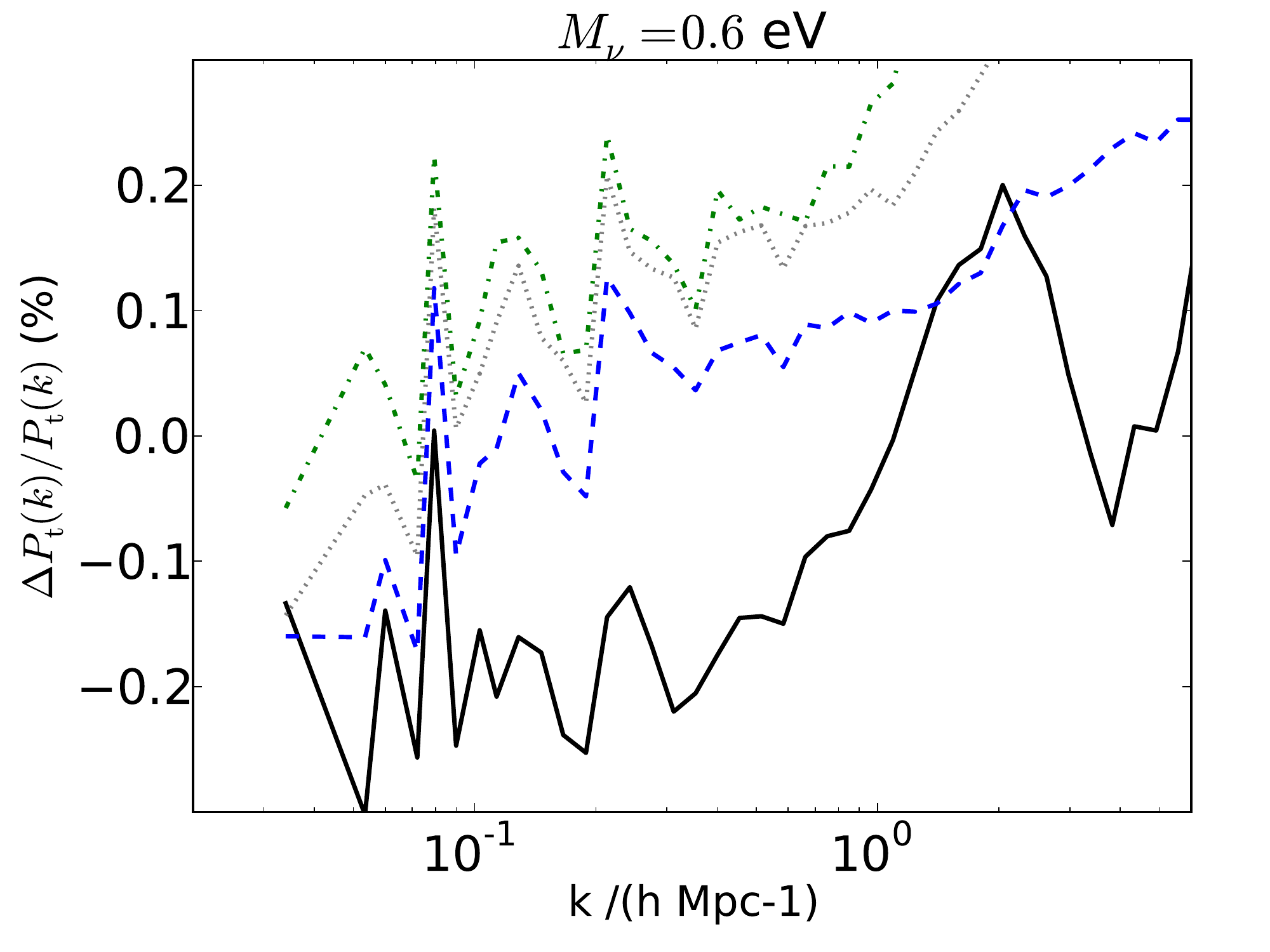}
\includegraphics[width=0.45\textwidth]{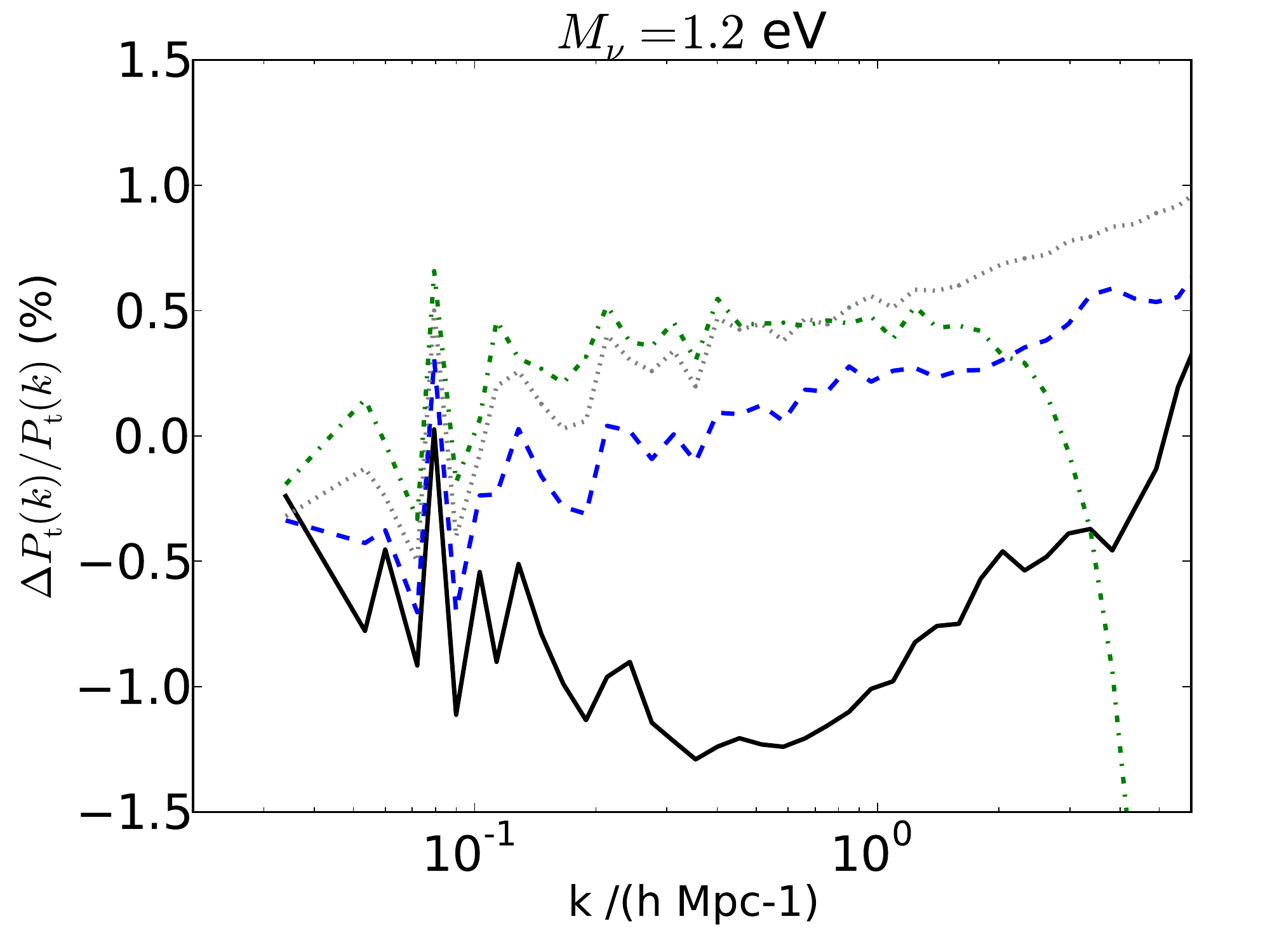}
\caption{Percentage differences in the total matter power spectrum between the 
semi-linear Fourier method and particles. $\Delta P_\mathrm{t} = P_\mathrm{Fourier} - P_\mathrm{Particle}$, so
positive values correspond to more power in the Fourier method. 
The simulations used were S05 (Top left), S10 (Top right), S20 (Bottom left) and S40 (Bottom right). 
Each line shows a different redshift: $z=9$ (green dot-dashed), $z=3$ (grey dotted), $z=1$ (blue dashed) 
and $z=0$ (black solid). 
}
\label{fig:P_tot_rel}
\end{figure*}


Figure \ref{fig:P_tot_z0} shows the suppression in the total matter power spectrum, 
defined as the ratio between the total matter power spectrum including
massive neutrinos with $M_\nu = \sum m_{\nu} = 0.3$ eV
and the matter power spectrum with massless neutrinos, but unchanged
$\Omega_{\rm M}$. The particle and Fourier neutrino simulation methods 
agree extremely well, and we see again the enhancement in the
suppression due to the delayed onset of non-linear growth, as
discussed in \cite{Bird_2012}. 

Figure \ref{fig:P_tot_rel} shows the percentage difference in the total power spectrum between the particle and semi-linear neutrino simulations, 
for redshifts between $z=3$ and $z=0$ and for a variety of neutrino masses. 
The agreement is extremely good in all cases. We show those scales where our simulation is resolved to less than $1\%$; note that 
the difference between the two methods is generally smaller than the
convergence error shown in Figures \ref{fig:P_tot_IC} and \ref{fig:P_tot_part}. 
The effect of non-linear growth in the neutrino component, neglected in our semi-linear method, only begins to become important 
for $M_\nu = 1.2$ eV, at $z < 1$. Since neutrinos of this mass are already ruled out by current data, 
this is unlikely to be a practical limitation.

One interesting feature is that on small scales we seem to see (slightly) more power in the Fourier method than the particle method, and the effect increases slightly at high redshift. This is not 
due to shot noise, which would lead to increased power in the particle method. This appears to be a convergence effect; as we 
have shown in Section \ref{sec:convergence}, these scales are not converged to less than $1\%$, and increasing the resolution of the 
simulation leads to a marginal increase of power on small scales. We therefore suspect that our semi-linear method has an 
effective resolution slightly higher than the particle method, although if we computed the Tree force for the neutrino particles, 
the situation would probably be reversed. 

It is interesting to compare our results to those obtained using the original fully-linear Fourier-space neutrino method of \cite{Brandbyge_2008}. Figure 1 of
that work is comparable to our Figure \ref{fig:P_tot_rel}. 
Our semi-linear method agrees with the particle-based method slightly better in the linear regime. 
This is due to our derivation of the neutrino power spectrum from the CDM, which incorporates sample variance in the neutrino component 
in a way very similar to the particle method.
In the non-linear regime, the agreement between the semi-linear method and the particle method is 
significantly better than that between the fully-linear and particle methods.
The fully-linear method shows differences from the particle method of $ >
1\%$ for neutrinos with $M_\nu = 0.6$ eV, and has an error of
about $5\%$ at $1.2$ eV. 
We ran test simulations using the fully-linear theory power spectrum,
but with our improved model for the neutrino phase structure
(i.e. assuming the neutrinos are always in phase with the CDM
rather than keeping their initial phases). The result was still in good agreement with \cite{Brandbyge_2008}, showing that the improved behaviour of the semi-linear 
method in the non-linear regime does indeed result from accounting for
the deeper non-linear potential wells sourcing the linear growth of neutrinos. 

\cite{Bird_2012}, using the linear theory neutrino implementation of \cite{Viel_2010}, 
found somewhat larger differences between the linear and particle methods than \cite{Brandbyge_2008}. 
When preparing this work, we discovered that this was due to a slight inconsistency in the algorithm used; 
the Fourier space neutrino implementation was altering the initial CDM transfer function to include neutrinos,
adding the neutrino power when calculating long-range forces, but neglecting them when outputting the power spectrum. 
Once this was corrected, we obtain fully-linear Fourier-space neutrino
results in good agreement with those of \cite{Brandbyge_2008}.

\subsection{Neutrino Power Spectra}

\begin{figure*}
\includegraphics[width=0.45\textwidth]{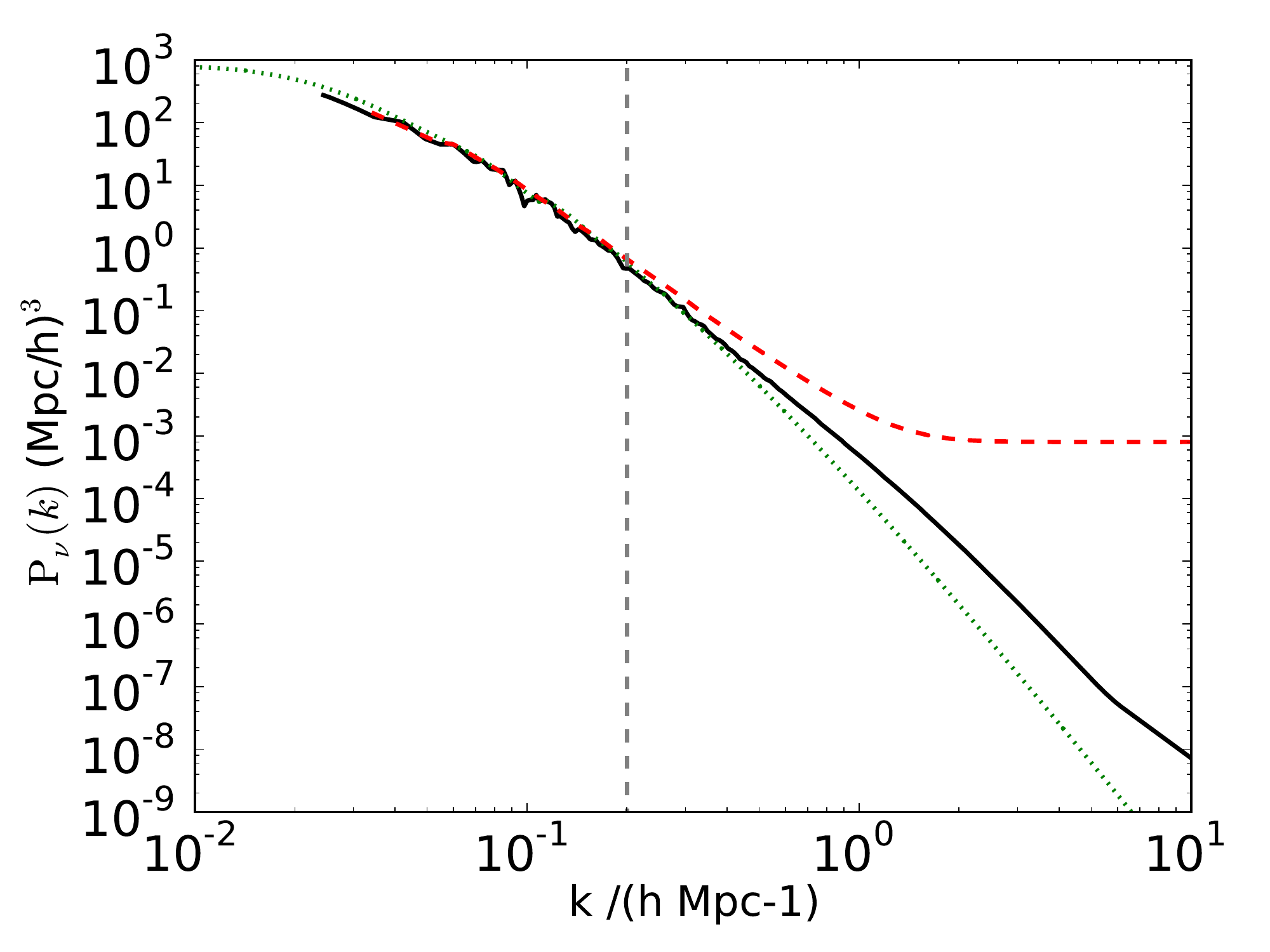}
\includegraphics[width=0.45\textwidth]{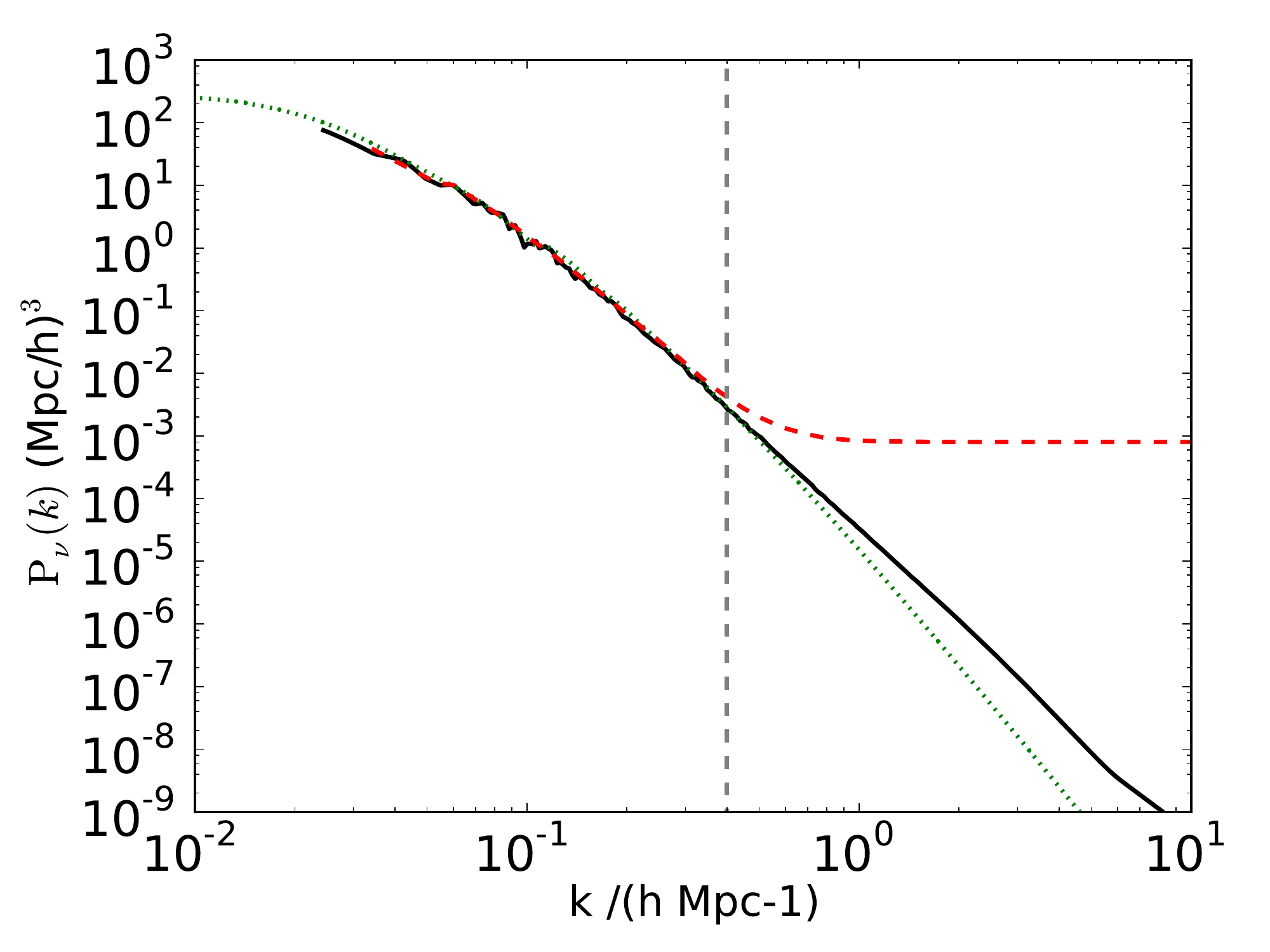}
\caption{The power spectrum of the neutrino component at (Left) $z=0$ (Right) $z=1$. 
Solid black shows the results of our semi-linear fourier-based method from simulation S10. 
Dashed red shows the particle method, from S10P, to obtain lower shot noise. 
Dotted green shows pure linear theory.
The vertical dashed grey line shows the approximate non-linear scale for the dark matter.
}
\label{fig:P_nu}
\end{figure*}

Figure \ref{fig:P_nu} shows the power spectra 
for the neutrino component at two different redshifts for the different methods. 
In the linear regime both codes agree well with linear theory. 
On non-linear scales our semi-linear method shows additional power
over the fully-linear neutrino power spectrum. 
On all scales, $P_\mathrm{CDM} \gg P_\nu$ and $k^3P_\nu/(2 \pi^2) \ll 1$. 

At $z=0$, the particle method produces additional power in the neutrino component. It begins to 
deviate from our semi-linear Fourier based method at $k = k_\mathrm{nl}$, the non-linear scale
for the dark matter. This first becomes apparent for $z < 0.5$, 
for all $M_\nu \geq 0.1$ eV.
We have checked that this area of the power spectrum is unchanged with a $512^3$ 
particle simulation, verifying that it is not being affected by shot noise. 
We ascribe this effect to non-linear growth in the neutrinos around the centres of clusters, as described in 
\cite{Ringwald_Wong_2004, Paco_2011}, and as we discussed in Section \ref{sec:clusters}.

This does not appear to have any effect on the total matter power spectrum. 
Most of the effect of neutrinos on the total matter power spectrum comes from their 
time integrated evolution; in particular the size of the trough is due to their effect on the 
onset of non-linear growth. Since the highly non-linear effects on the 
neutrino power spectrum only take place at relatively late times, 
the impact on the total power spectrum is minimal. Moreover, on scales
much smaller than the free-streaming scale, the main effect of
including neutrinos is essentially to reduce the matter overdensity by
a factor $(1 - f_{\nu})$, while keeping the same background expansion
rate \citep{Lesgourgues_2006}. As long as $|\delta_{\nu}| \ll
|\delta_{\rm cdm}|$, the exact value of the neutrino overdensity does
not matter very much. The same conclusion was reached by
\cite{Shoji_Komatsu_2009}, who compared their thrid-order perturbation
theory calculations (for both CDM and neutrinos, the latter approximated as a
perfect fluid with pressure) to a computation where CDM is treated
to third order but neutrinos are computed to linear order (without
accounting for non-linear growth of potentials), similar to the
treatment of \cite{Saito_2008}. They found that
although neutrino overdensities can be underestimated by order unity
on non-linear scales, the total matter power spectrum is very accurately obtained, even with
the simpler method\footnote{In the notation of
  \cite{Shoji_Komatsu_2009}, our method assumes $P_{\rm tot} = f_c^2
  P_{\infty, c} + (2 f_c f_{\nu} g_1 + f_{\nu}^2 g_1^2)P_{\infty, c}$,
which is better than the treatment of \cite{Saito_2008}, in the sense
that we use the full non-linear CDM power spectrum as a source for
neutrino overdensities.}.

Finally, let us point out that for $z > 0.5$, subtracting the scale-free shot noise from $P_\nu$ in the particle simulation produces very good agreement 
with the semi-linear Fourier method, even at scales where the shot noise dominates. This is further evidence that neutrino 
shot noise is not having a strong dynamical effect, and is not causing spurious clustering.

\subsection{Performance}

Simulations S05-S20 were consistently faster when using our Fourier method. The speed increase was $13\%$ of the total 
walltime (which includes time spent reading and writing to disc). Note that the slowest single algorithm in \gadget~is 
the Tree method for computing short-range forces, which is disabled for neutrinos even for our particle based simulations, 
hence a large proportion of the execution time is independent of the method used to simulate neutrinos. 
More importantly, the total memory usage of \gadget was $40\%$ smaller in the Fourier method than with particle neutrinos, 
essentially identical to the memory usage of a pure dark matter simulation. 
This is important because memory is often the limiting factor when performing large modern simulations. 

The S10P simulation, which had $8$ times more dark matter particles than S10, took $12$ times longer. 
This scaling is similar to that expected for a pure dark matter simulation, demonstrating that our neutrino method scales well.
In fact, the only limit to scalability in the neutrino calculation is the need for inter-process communication when computing the power spectrum.

Overall, our Fourier method appears to have similar performance characteristics to a pure dark matter simulation, as should be expected; 
the time to compute the neutrino power spectrum is completely negligible compared to the $N$-body algorithms, and the most costly part of 
our Fourier algorithm is summing modes on the Fourier-transformed density grid to compute the power spectrum.

\section{Applications} \label{sec:applications}

\subsection[Lyman-alpha Forest]{\Lya Forest}

\begin{figure}
\includegraphics[width=0.45\textwidth]{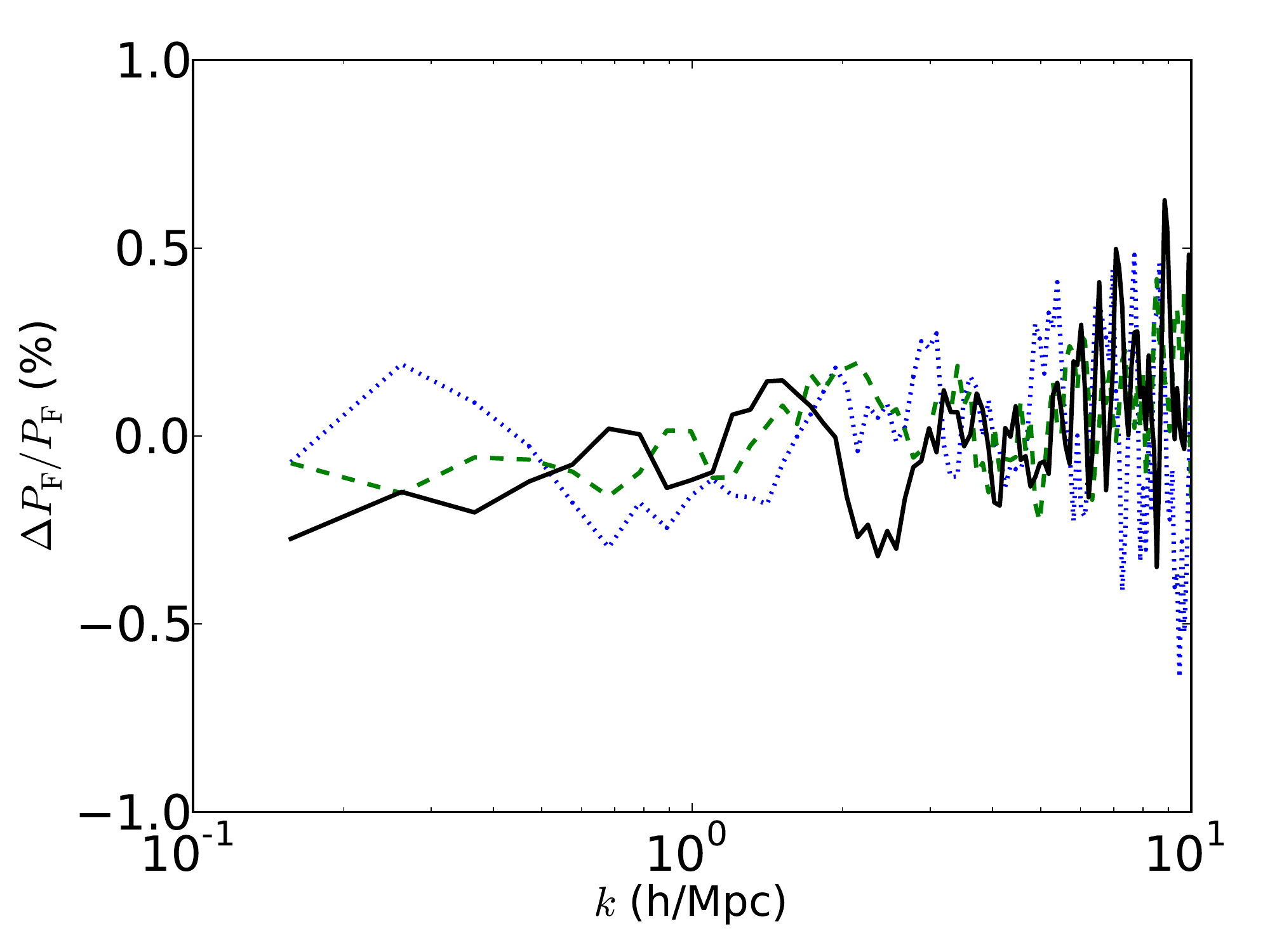}
\caption{The change in the flux power spectrum between our Fourier method 
and the particle method, using simulations F10BA with $m_\nu = 0.1$ eV.
Positive values correspond to more power with the Fourier method. 
Each line shows a different redshift $z=4$ (blue dotted), $z=3$ (green dashed), and $z=2$ (black solid).
SDSS \Lya data constrains the flux power spectrum for $k \lesssim 4 \hMpc$ \protect\cite{McDonald_2004data}. 
}
\label{fig:P_flux}
\end{figure}

The \Lya forest is an indirect probe of the matter power spectrum at small, non-linear 
scales ($k = 0.1 - 4 \hMpc$), and at high redshift, $z=2-4$. 
The power spectrum of the \Lya flux measures the clustering of the absorption signal 
from neutral hydrogen in quasar spectra, and can be used to place constraints on the 
amplitude of primordial perturbations. When combined with constraints from large scales, this can lead to tight constraints 
on neutrino mass \citep{Seljak_2000,Gratton_2007,Viel_2010}. At these high redshifts, 
shot noise could be an issue for light neutrinos, so it is a natural place to apply our method. 
In addition, simulations with neutrino particles, dark matter and baryons can become unwieldy. 

We used simulation F10BA to simulate the \Lya forest at $z=2-4$. 
Since the \Lya forest is not the main focus of our paper, we shall not explain in detail our 
simulation methodology, and refer the interested reader to \cite{Viel_2010}.
Our simulation calculates the clustering and ionisation state of the gas at $z=2$ using smooth particle hydrodynamics, together 
with radiative cooling and a reaction network which assumes optically thin gas in ionisation equilibrium.
We then simulate the flux in a set of $16000$ quasar spectra by calculating the absorption along random 
lines of sight. The averaged power spectrum of this flux is the observable quantity. Although constraints are available 
for $k \lesssim 4 \hMpc$, the \Lya forest is also sensitive to the collapse scale of hydrogen absorbers $k \sim 60 \hMpc$, 
so very high resolution simulations are required. Figure \ref{fig:P_flux} shows the change in the flux power spectrum between our two methods; 
clearly the differences are quite small. For simulations with the resolution of F10BA and $m_\nu = 0.1$ eV, therefore, shot noise 
is not having a significant affect on the dynamics, even at $z=4$, as also found by \cite{Viel_2010}.
The total effect of neutrinos on the flux power is $5-10\%$.

\subsection{Hierarchy}

\begin{figure}
\includegraphics[width=0.45\textwidth]{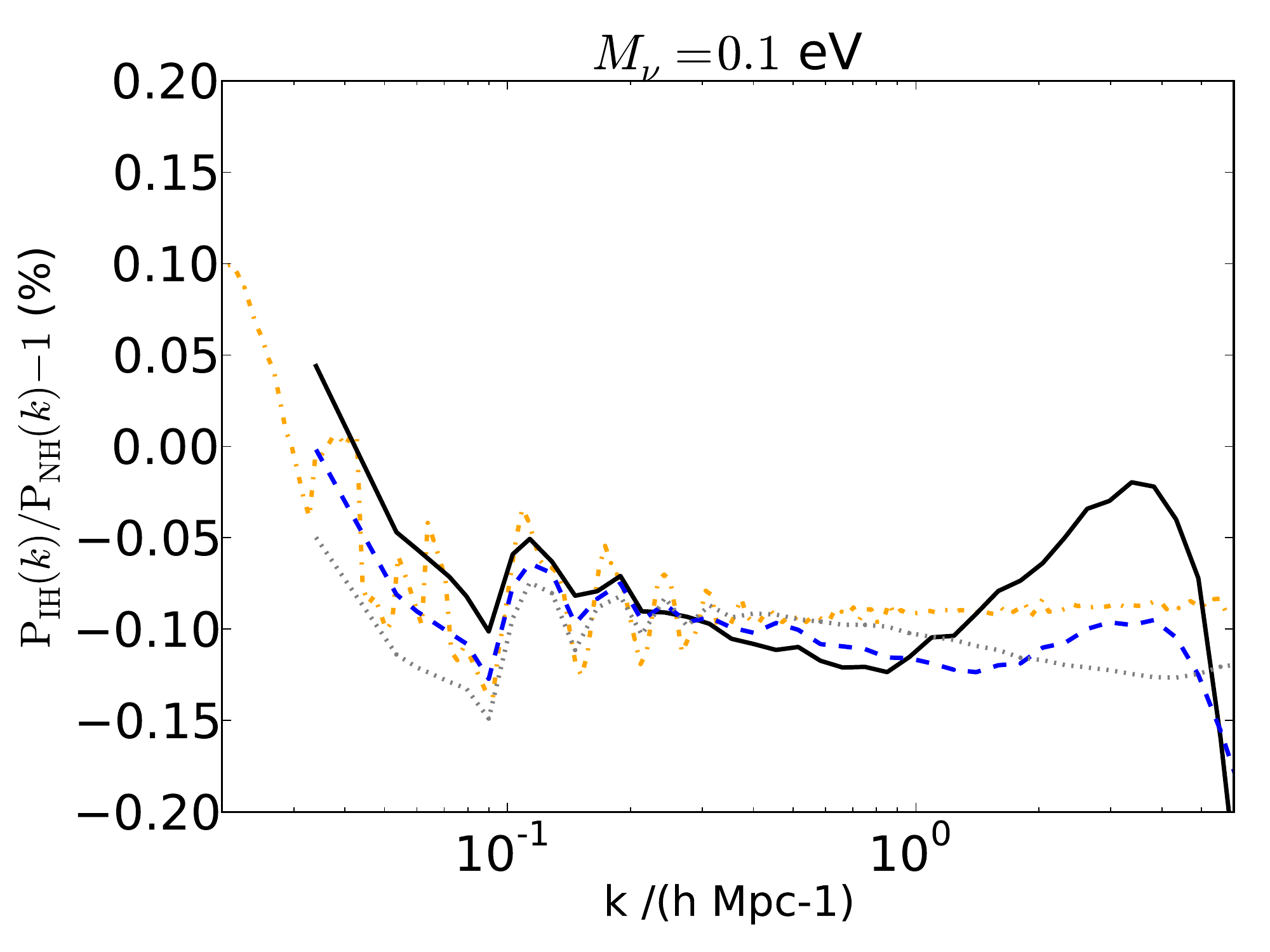}
\caption{The difference in the total matter power spectrum between the 
inverted and normal neutrino hierarchies.
Negative values correspond to more power in the normal scenario. 
Each line shows a different redshift: $z=3$ (grey dotted), $z=1$ (blue dashed) and $z=0$ (black solid). 
The orange dot-dashed line shows the linear prediction at $z=0$.
}
\label{fig:P_tot_hier}
\end{figure}

Achieving maximal accuracy for small neutrino masses requires us to account for the neutrino hierarchy. 
This is difficult in particle-based codes, as two neutrino species doubles the amount of memory required,
but has been done \citep{Wagner_2012}. It is, however, essentially trivial using our method. We have 
performed two simulations (S03IH and S03NH) with a total neutrino mass
of $M_\nu = 0.1$ eV using either the normal or inverted hierarchies, both to demonstrate the technique and examine the effects of the hierarchy.
For completeness, the masses of the three neutrino species were
$0.022$, $0.024$, $0.054$~eV for S03NH 
and $0$, $0.049$, $0.051$~eV for S03IH.
Our results are shown in Figure \ref{fig:P_tot_hier}, together with the prediction of linear theory. 
The linear theory change due to the hierarchy is quite small (below our convergence error). 
Furthermore, the effect in linear theory is mostly due to the change in background evolution 
caused by one of the possible hierarchies having one massless species.
As shown in Figure 16 of \cite{Lesgourgues_2006}, the difference between 
the hierarchies when all three neutrino species are massive in both cases is much less.
Although differences in our cosmological parameters mean that the linear effect is smaller 
than that found by \cite{Wagner_2012}, we still find, as they did, an enhancement at non-linear scales, 
analogous to the enhancement for the total neutrino effect shown in Figure \ref{fig:P_tot_z0}.
Determining the hierarchy separately from the total neutrino mass from an effect this subtle
is likely to prove challenging. However, obtaining robust constraints with $M_\nu \lesssim 0.1$eV 
will require accounting for the mass splitting.

\subsection{21cm Forest}

Observations of the spin-flip transition of neutral hydrogen in the $21$cm forest have the potential to probe 
structure growth at $z\sim 20$ \citep{McQuinn_2006}. Structure formation at these redshifts is sensitive to the non-linear growth 
of the first, small halos. No-one has yet examined the impact of neutrinos on the $21$cm forest, and performing the 
requisite simulations is beyond the scope of this paper. However, our code would be ideal for such a study. 
Halos at these redshifts are not yet sufficiently massive for our method to break down in their central region, 
so one could probe arbitrarily small scales. 
Furthermore, shot noise would be a severe issue for the particle method at these redshifts. 

\section{Discussion and Conclusions} \label{sec:conclusion}

We have presented a new semi-linear method for simulating the effects of massive neutrinos on 
cosmic structure, taking advantage of the hierarchy between the neutrino
free-streaming scale and the non-linear scale, $k_{\rm fs} < k_{\rm
  nl}$ for currently allowed neutrino masses. The power spectrum of the neutrino component is calculated using 
perturbation theory, and added to the cold dark matter in Fourier-space, as in 
\cite{Brandbyge_2008}. However, we improve upon their method by 
sourcing neutrino perturbations using the gravitational potential
obtained from the full non-linear dark matter overdensity rather than from
the dark matter power predicted by linear theory. We evolve the CDM,
baryons and neutrinos simultaneously and self-consistently.

For observationally relevant neutrino masses, our method gives 
results for the non-linear power spectrum essentially identical to
a fully converged neutrino particle treatment, a significant increase in accuracy 
over fully linear Fourier-space neutrinos. 
Our method also has several advantages over a particle treatment; it is faster, 
uses significantly less memory, and does not suffer from shot noise in the 
neutrino component. Furthermore, it allows for an easy inclusion of
the correct background evolution with relativistic corrections, and makes inclusion of the neutrino hierarchy trivial. 
These properties make it especially suitable for simulating neutrinos at
the low end of the currently allowed mass range.

Our treatment is not strictly valid in the inner regions of very 
massive halos, where the escape velocity might be larger than the thermal velocity of
neutrinos, which may be captured and cluster non-linearly. Accurately describing the distribution of neutrinos in these halos requires 
treating them as $N$-body particles. Because of this effect, 
we do not accurately recover the non-linear neutrino power spectrum at 
redshifts $z < 0.5$. However, this does not affect the total matter 
power spectrum, because on the one hand most of the neutrino effect is imprinted
at earlier times, where our method is extremely accurate, and on the
other hand, even with an enhanced clustering due to non-linear growth,
neutrino overdensities are very subdominant compared to those of
the cold dark matter.

While we have focussed on neutrinos, our treatment can also be easily 
applied to any hot dark matter particle whose characteristic velocity 
is larger than (or at least of the order of) the escape velocity of
massive clusters, and whose free-streaming length today is larger than
the non-linear scale. 
There are many applications of our method beyond the non-linear matter power spectrum; 
we have briefly discussed the \Lya and 21cm forests. 
It allows the inclusion of neutrinos in an $N$-body simulation at minimal cost 
and effort, and should thus be useful for any problem in 
non-linear structure formation. 

\section*{Acknowledgements}
We thank Volker Springel for allowing one of us (S.~B.) to use 
the code \textsc{Gadget-3}, Matteo Viel for useful discussions and 
Daniel Grin for enlightening conversations on
various aspects of this work. We are grateful to Anthony Challinor,
Julien Lesgourgues and Eiichiro Komatsu for
providing detailed comments on the draft of this paper, and
acknowledge conversations with Matias Zaldarriaga, Scott Tremaine, Marilena LoVerde and
Tobias Heinemann. The authors are supported by the National Science
Foundation grants AST-080744 (for Y. A.-H.) and AST-0907969
(for S.~B.). 

\appendix

\section{Analytic approximation for linear HDM clustering}

In this appendix we derive an approximate expression for $\delta_{\rm
  hdm}$ valid on all scales during matter domination. We do not use this expression when
evaluating neutrino clustering, but it provides some insight. Our main simplification here
is to assume that the gravitational
potential is nearly constant in time (or equivalently that the matter
overdensity grows roughly linearly with the scale factor). During matter domination and for linear evolution,
$\phi$ is in fact strictly constant. With this approximation, we get
\barr
&&\delta_{\rm hdm}(\bs k, \tau) \approx \delta^I_{\rm hdm}(\bs k, \tau_i, \tau)\nonumber\\
&&~- k^2 \phi(\bs k, \tau) \int_{0}^{s-s_i}
\mathcal{I}\left[\frac{k}{m}\tilde{s}\right] [a(s-\tilde{s})]^2 \tilde{s}~
d\tilde{s}, 
\earr
where we made the change of variables $\tilde{s} \equiv s -
s'$. Assuming matter domination, we have $a \propto \tau^2$, $\tau =
2/(a H)$, and 
\beq
a(s - \tilde s) = \frac{a(s)}{\left(1 + \frac{1}{2} a^2(s) H(s) \tilde{s} \right)^2}.
\eeq
Changing variables to $X = (k/m) \tilde{s}$, we obtain
\barr
&&\delta_{\rm hdm}(\bs k, \tau) \approx \delta^I_{\rm hdm}(\bs k, \tau_i, \tau)\nonumber\\
&&~- (m a)^2 \phi(\bs k, \tau) \int_{0}^{\frac{k(s-s_i)}m}
\frac{X \mathcal{I}(X)}{\left( 1 + X/X_k \right)^4} d X, 
\earr
where $X_k \equiv 2 k/(m a^2 H) \sim q_0^{-1} k/k_{\rm fs}(a)$. Let us now consider scales small
enough that initial conditions are ``forgotten'', $k q_0(s - s_i)/m
\gg 1$. In this case $\delta^{I} \approx 0$ and the upper limit of the integral in the above
equation can be approximated as infinity:
\beq
\delta_{\rm hdm}(\bs k, \tau) \approx - (m a)^2 \phi(\bs k, \tau) \int_{0}^{+\infty}
\frac{X \mathcal{I}(X)}{\left( 1 + X/X_k \right)^4} d X.
\eeq
Note that during matter domination, 
\beq
q_0 X_k \approx \sqrt{\frac{a_i}{a}} \frac{k q_0(s - s_i)}{m},
\eeq
so provided $a/a_i$ is sufficiently large, $q_0 X_k$ may be of order
unity, even if $k q_0(s - s_i)/m \gg 1$. Using Poisson's equation
(\ref{eq:Poisson}), we arrive at
\beq
\delta_{\rm hdm}(\bs k, \tau) \approx \mathcal{F}(k/k_{\rm fs})
\delta_{\rm M}(\bs k, \tau),
\eeq
where we defined the dimensionless function
\beq
\mathcal{F}(k/k_{\rm fs}) \equiv \frac{6}{X_k^2}\int_{0}^{+\infty}
\frac{X \mathcal{I}(X)}{\left( 1 + X/X_k \right)^4} d X.
\eeq
For $k \ll k_{\rm fs}$, $q_0 X_k \rightarrow 0$ and
$\mathcal{F}(k \ll k_{\rm fs})\rightarrow 1$, i.e. on large scales,
$\delta_{\rm hdm} \approx \delta_{\rm M}$. For $k \gg k_{\rm fs}$, $q_0 X_k \rightarrow \infty$ and one can show
that
\beq
\mathcal{F}(k \gg k_{\rm fs}) = \frac6{X_k^2}\overline{q^{-2}},
\eeq
where
\beq
\overline{q^{-2}} \equiv \frac{\int dq f_0(q)}{\int dq q^2 f_0(q)}.
\eeq
In this limit we recover the small-scale solution of Section
\ref{sec:small-scales}, $\delta_{\rm hdm}  = - \overline{v^{-2}} \phi = \left(k_{\rm fs}/k\right)^2 \delta_{\rm M}$.

We evaluated the function $\mathcal{F}$ numerically for massive
neutrinos and found that it is approximated to better than
12\% for any value of $k/k_{\rm fs}$ by the very simple form (see also \cite{Wong_2008})
\beq
\mathcal{F}(k/k_{\rm fs}) \approx \frac{1}{\left(1 + k/k_{\rm fs}\right)^2}.
\eeq

\section{Accuracy of the iterative solution for the neutrino power spectrum}\label{app:iteration}
In this appendix we discuss the accuracy of the step where we obtain
the neutrino power spectrum at time $\tau$ given the total power
spectrum at earlier times $P_{\rm M}(\tau' \leq \tau - \Delta \tau)$,
and the current CDM power spectrum, $P_{\rm cdm}(\tau)$. We start by
noticing that, because of our assumption of total correlation of
neutrinos and CDM [Eq.~(\ref{eq:phases})], we have
\beq
P_{\rm M}^{1/2}(k, \tau) = (1 - f_{\nu}) P_{\rm cdm}^{1/2}(k, \tau) +
f_{\nu} P_{\nu}^{1/2}(k, \tau).
\eeq
We now split Eq.~(\ref{eq:P-final}) in a term that is completely known
at the current time step and a term that depends on the neutrino power
spectrum between $\tau - \Delta \tau$ and $\tau$
\beq
P_{\nu}^{1/2}(k, \tau) = P_{\nu}^{1/2}(k, \tau; \textrm{known}) +
\Delta P_{\nu}^{1/2}(\textrm{implicit}), 
\eeq
where
\beq
\Delta P_{\nu}^{1/2}(\textrm{implicit}) \equiv \frac32 \Omega_{\rm M} H_0^2 f_{\nu} \int_{\tau - \Delta \tau}^{\tau}  (s - s')\mathcal{I}_{s', s}
P^{1/2}_{\nu}(\tau')d \tau'.
\eeq
We see that this renders the equation for $P_{\nu}(k, \tau)$
implicit. Now, provided $H \Delta t = a H \Delta \tau \ll 1$, which is
obviously the case in a $N$-body code, the neutrino power spectrum in
the integral is approximately $P_{\nu}(\tau') = P_{\nu}(\tau)[1 + \mathcal{O}(H
\Delta t)]$. Switching to the integration variable $s'$, and using
$d \tau' = a(\tau') ds' \approx a(\tau) ds'$, we obtain, up to
corrections of \emph{relative} order $\mathcal{O}(H \Delta t)$,
\beq
\Delta P_{\nu}^{1/2}(\textrm{implicit}) = \epsilon(\tau)
P_{\nu}^{1/2}(k, \tau), 
\eeq
where we have defined
\beq
\epsilon(\tau) \equiv \frac32 \Omega_{\rm M} H_0^2
f_{\nu} a(\tau) \int_{s-\Delta s}^{s} (s - s')
\mathcal{I}_{s',s} d s',
\eeq
where $\Delta s \approx \Delta \tau/a \approx \Delta t/a^2$.
Now $\mathcal{I} \leq 1$ for any value of its argument, and therefore
\barr
\epsilon(\tau) &\leq& \frac34 \Omega_{\rm M} H_0^2 a^{-3}
f_{\nu} (\Delta t)^2 = \frac34 f_{\nu} \Omega_{\rm M}(a) (H \Delta
t)^2\nonumber\\
 &<& f_{\nu} (H
\Delta t)^2.
\earr
Therefore, whereas the exact solution is formally
\beq
P_{\nu}^{1/2}(k, \tau) = \frac{1}{1+ \epsilon(k, \tau)} P_{\nu}^{1/2}(k, \tau; \textrm{known}),
\eeq
because $\epsilon = \mathcal{O}\left(f_{\nu}(H \Delta t)^2\right)$ is
small, an iterative solution will converge very
rapidly, the error after $n$ iterations being of the order of
$\epsilon^n$ [where the $n=0$ iteration is $P_{\nu} = 0$ and the $(n+1)$-th
iteration is obtained by using $P_{\nu}(\rm implicit)$ obtained from
the $n$-th iteration]. Instead of initialising $P_{\nu}$ with 0, we
moreover initialise it assuming $P_{\nu}(\tau) \approx P_{\nu}(\tau -
\Delta \tau)$ in the integral. The overall error is therefore
extremely small, of the
order of $\mathcal{O}\left(f_{\nu}(H \Delta t)^3\right)$.

\section[Fitting Function for I(X)]{Fitting Function for $\mathcal{I}(X)$}
\label{sec:fitIX}

\begin{figure}
\includegraphics[width = 80 mm]{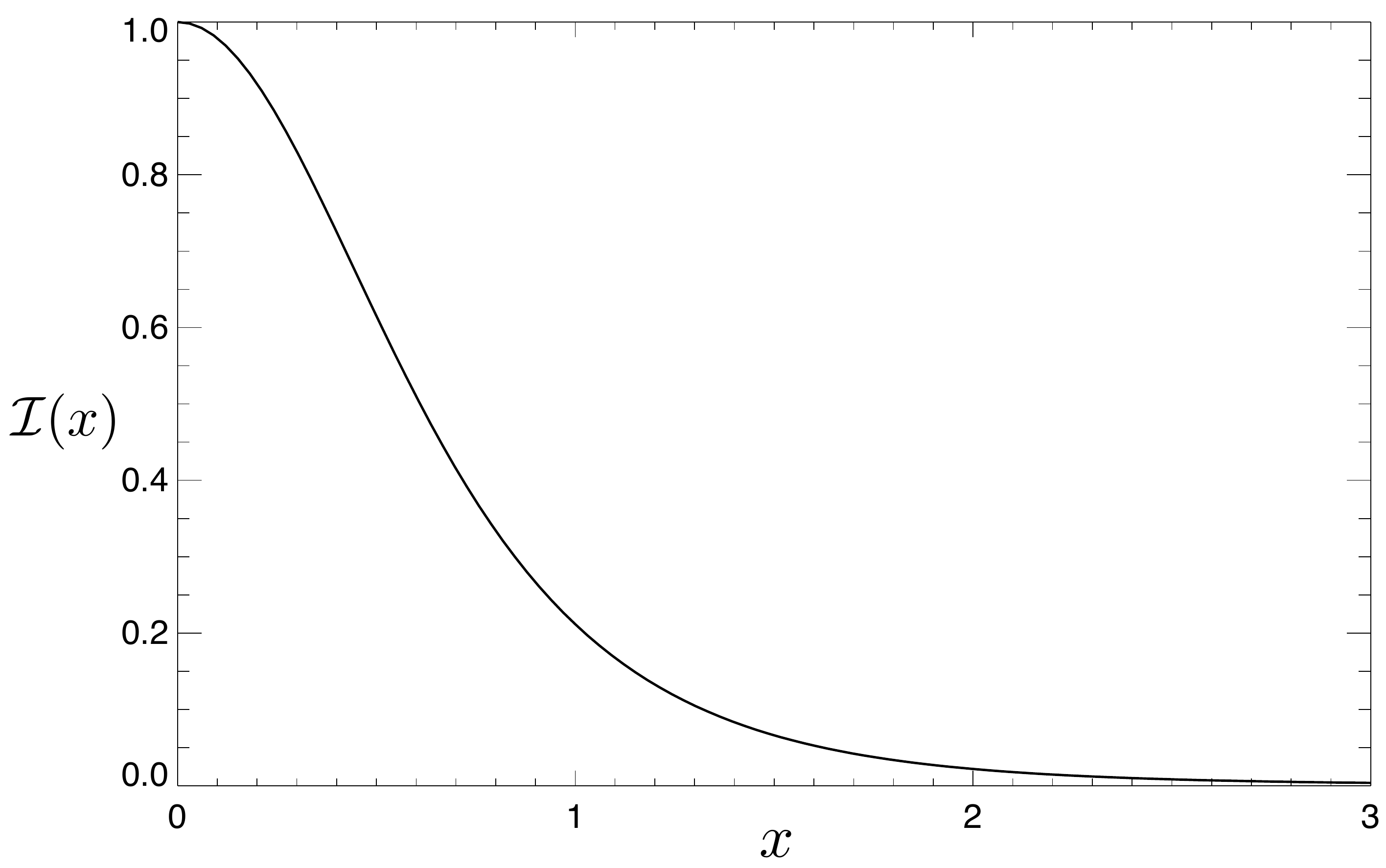}
\caption{Function $\mathcal{I}(x)$ that determines the damping of perturbations due to
  free-streaming of neutrinos [Eq.~(\ref{eq:I.def}) with $x = X T_{\nu,0}$]} \label{fig:specialJ} 
\end{figure}

The function $\mathcal{I}(X)$ only depends on the dimensionless product $x \equiv X T_{\nu,0}$ (we remind the
reader that $X$ was defined as an inverse comoving momentum). For
neutrinos described with the relativistic Fermi-Dirac distribution, it can
obtained from the following series expansion \citep{Bertschinger_Watts_1988}:
\beq
\mathcal{I}(x) = \frac{4}{3 \zeta(3)}\sum_{n=1}^{\infty}(-1)^{n+1}
\frac{n}{(n^2 + x^2)^2}.
\eeq
We also find that $\mathcal{I}(x)$ is approximated to very high
accuracy by the fit
\beq
\mathcal{I}_{\rm fit}(x) \equiv \frac{1 + 0.0168 ~x^2 + 0.0407 ~x^4}{1 + 2.1734
 ~ x^2 + 1.6787 ~x^{4.1811} + 0.1467~ x^8}. \label{eq:fit}
\eeq
This fit gives the correct asymptotic behaviours
\barr
\mathcal{I}(x) &\approx& 1 - \frac{5 \zeta(5)}{2 \zeta(3)} x^2 \approx
1 - 2.1566 ~x^2 , \ \ \ x \ll 1,\\
\mathcal{I}(x) &\approx& \frac{1}{3 \zeta(3) x^4} 
\approx \frac{0.2773}{x^4},  \ \ \ \ \ \ \ \ \ \ \ \ \ \ \ \ \ x \gg 1.
\earr
The relative accuracy of the fit is $|\Delta \mathcal{I}/\mathcal{I}| \lesssim
1\%$ for $0 \leq x \leq 2$ and $|\Delta \mathcal{I}/\mathcal{I}| \lesssim
3 \%$ for $0 \leq x \leq +\infty$; the absolute accuracy is $|\Delta
\mathcal{I}| \lesssim 0.07\%$ for $0 \leq x \leq +\infty$. We show the function $\mathcal{I}$ in Fig.~\ref{fig:specialJ}.
We have checked explicitly that modifying our code to use the full function instead of the fitting formula does not affect our results.

\label{lastpage}

\bibliography{neutrinos}

\begin{thebibliography}{}

\bibitem[\protect\citeauthoryear{{Abazajian}, {Switzer}, {Dodelson}, {Heitmann}
  \& {Habib}}{{Abazajian} et~al.}{2005}]{Abazajian_Switzer_2005}
{Abazajian} K.,  {Switzer} E.~R.,  {Dodelson} S.,  {Heitmann} K.,    {Habib}
  S.,  2005, \prd, 71, 043507, \eprint{astro-ph/0411552},
  \adsurl{http://adsabs.harvard.edu/abs/2005PhRvD..71d3507A}

\bibitem[\protect\citeauthoryear{{Abazajian} et~al.,}{{Abazajian}
  et~al.}{2011}]{Abazajian}
{Abazajian} K.~N.,  et~al., 2011, Astroparticle Physics, 35, 177,
  \eprint{1103.5083},
  \adsurl{http://adsabs.harvard.edu/abs/2011APh....35..177A}

\bibitem[\protect\citeauthoryear{{Beaulieu} et~al.,}{{Beaulieu}
  et~al.}{2010}]{Euclid_2010}
{Beaulieu} J.~P.,  et~al., 2010, in Pathways Towards Habitable Planets Vol.~430
  of ASP Conf. Ser., {EUCLID: Dark Universe Probe and Microlensing Planet
  Hunter}.
pp 266--+, \eprint{1001.3349},
  \adsurl{http://adsabs.harvard.edu/abs/2010ASPC..430..266B}

\bibitem[\protect\citeauthoryear{{Bertschinger} \& {Watts}}{{Bertschinger} \&
  {Watts}}{1988}]{Bertschinger_Watts_1988}
{Bertschinger} E.,  {Watts} P.~N.,  1988, ApJ, 328, 23,
  \adsurl{http://adsabs.harvard.edu/abs/1988ApJ...328...23B}

\bibitem[\protect\citeauthoryear{{Bird}, {Viel} \& {Haehnelt}}{{Bird}
  et~al.}{2012}]{Bird_2012}
{Bird} S.,  {Viel} M.,    {Haehnelt} M.~G.,  2012, MNRAS, p.~2175,
  \eprint{1109.4416},
  \adsurl{http://adsabs.harvard.edu/abs/2012MNRAS.tmp.2175B}

\bibitem[\protect\citeauthoryear{{Bond}, {Efstathiou} \& {Silk}}{{Bond}
  et~al.}{1980}]{Bond_1980}
{Bond} J.~R.,  {Efstathiou} G.,    {Silk} J.,  1980, Physical Review Letters,
  45, 1980, \adsurl{http://adsabs.harvard.edu/abs/1980PhRvL..45.1980B}

\bibitem[\protect\citeauthoryear{{Bond} \& {Szalay}}{{Bond} \&
  {Szalay}}{1983}]{Bond_Szalay_1983}
{Bond} J.~R.,  {Szalay} A.~S.,  1983, ApJ, 274, 443,
  \adsurl{http://adsabs.harvard.edu/abs/1983ApJ...274..443B}

\bibitem[\protect\citeauthoryear{{Brandbyge} \& {Hannestad}}{{Brandbyge} \&
  {Hannestad}}{2009}]{Brandbyge_2009}
{Brandbyge} J.,  {Hannestad} S.,  2009, \jcap, 5, 2, \eprint{0812.3149},
  \adsurl{http://adsabs.harvard.edu/abs/2009JCAP...05..002B}

\bibitem[\protect\citeauthoryear{{Brandbyge} \& {Hannestad}}{{Brandbyge} \&
  {Hannestad}}{2010}]{Brandbyge_2010}
{Brandbyge} J.,  {Hannestad} S.,  2010, \jcap, 1, 21, \eprint{0908.1969},
  \adsurl{http://adsabs.harvard.edu/abs/2010JCAP...01..021B}

\bibitem[\protect\citeauthoryear{{Brandbyge}, {Hannestad}, {Haugb{\o}lle} \&
  {Thomsen}}{{Brandbyge} et~al.}{2008}]{Brandbyge_2008}
{Brandbyge} J.,  {Hannestad} S.,  {Haugb{\o}lle} T.,    {Thomsen} B.,  2008,
  \jcap, 8, 20, \eprint{0802.3700},
  \adsurl{http://adsabs.harvard.edu/abs/2008JCAP...08..020B}

\bibitem[\protect\citeauthoryear{{Brandbyge}, {Hannestad}, {Haugb{\o}lle} \&
  {Wong}}{{Brandbyge} et~al.}{2010}]{Brandbyge_2010a}
{Brandbyge} J.,  {Hannestad} S.,  {Haugb{\o}lle} T.,    {Wong} Y.~Y.~Y.,  2010,
  \jcap, 9, 14, \eprint{1004.4105},
  \adsurl{http://adsabs.harvard.edu/abs/2010JCAP...09..014B}

\bibitem[\protect\citeauthoryear{{Brandenberger}, {Kaiser} \&
  {Turok}}{{Brandenberger} et~al.}{1987}]{Brandenberger_1987}
{Brandenberger} R.,  {Kaiser} N.,    {Turok} N.,  1987, \prd, 36, 2242,
  \adsurl{http://adsabs.harvard.edu/abs/1987PhRvD..36.2242B}

\bibitem[\protect\citeauthoryear{{Carbone}, {Verde}, {Wang} \&
  {Cimatti}}{{Carbone} et~al.}{2011}]{Carbone_2011}
{Carbone} C.,  {Verde} L.,  {Wang} Y.,    {Cimatti} A.,  2011, \jcap, 3, 30,
  \eprint{1012.2868},
  \adsurl{http://adsabs.harvard.edu/abs/2011JCAP...03..030C}

\bibitem[\protect\citeauthoryear{{Cooray}}{{Cooray}}{1999}]{Cooray_1999}
{Cooray} A.~R.,  1999, A\&A, 348, 31, \eprint{astro-ph/9904246},
  \adsurl{http://adsabs.harvard.edu/abs/1999AA...348...31C}

\bibitem[\protect\citeauthoryear{{de Putter} et~al.,}{{de Putter}
  et~al.}{2012}]{SDSS_2012}
{de Putter} R.,  et~al., 2012, ArXiv e-prints, \eprint{1201.1909},
  \adsurl{http://adsabs.harvard.edu/abs/2012arXiv1201.1909D}

\bibitem[\protect\citeauthoryear{{de Vega} \& {Sanchez}}{{de Vega} \&
  {Sanchez}}{2012a}]{Vega1}
{de Vega} H.~J.,  {Sanchez} N.~G.,  2012a, \prd, 85, 043516,
  \eprint{1111.0290},
  \adsurl{http://adsabs.harvard.edu/abs/2012PhRvD..85d3516D}

\bibitem[\protect\citeauthoryear{{de Vega} \& {Sanchez}}{{de Vega} \&
  {Sanchez}}{2012b}]{Vega2}
{de Vega} H.~J.,  {Sanchez} N.~G.,  2012b, \prd, 85, 043517,
  \eprint{1111.0300},
  \adsurl{http://adsabs.harvard.edu/abs/2012PhRvD..85d3517D}

\bibitem[\protect\citeauthoryear{{Dehnen} \& {Read}}{{Dehnen} \&
  {Read}}{2011}]{Nbody}
{Dehnen} W.,  {Read} J.~I.,  2011, European Physical Journal Plus, 126, 55,
  \eprint{1105.1082},
  \adsurl{http://adsabs.harvard.edu/abs/2011EPJP..126...55D}

\bibitem[\protect\citeauthoryear{{Eitel}}{{Eitel}}{2005}]{Eitel_2005}
{Eitel} K.,  2005, Nuclear Physics B Proceedings Supplements, 143, 197,
  \adsurl{http://adsabs.harvard.edu/abs/2005NuPhS.143..197E}

\bibitem[\protect\citeauthoryear{{Fogli}, {Lisi}, {Marrone}, {Montanino},
  {Palazzo} \& {Rotunno}}{{Fogli} et~al.}{2012}]{Fogli_2012}
{Fogli} G.~L.,  {Lisi} E.,  {Marrone} A.,  {Montanino} D.,  {Palazzo} A.,
  {Rotunno} A.~M.,  2012, \prd, 86, 013012, \eprint{1205.5254},
  \adsurl{http://adsabs.harvard.edu/abs/2012PhRvD..86a3012F}

\bibitem[\protect\citeauthoryear{{Gilbert}}{{Gilbert}}{1966}]{Gilbert_1966}
{Gilbert} I.~H.,  1966, ApJ, 144, 233,
  \adsurl{http://adsabs.harvard.edu/abs/1966ApJ...144..233G}

\bibitem[\protect\citeauthoryear{{Gnedin} \& {Hui}}{{Gnedin} \&
  {Hui}}{1998}]{Gnedin_1998}
{Gnedin} N.~Y.,  {Hui} L.,  1998, MNRAS, 296, 44, \eprint{astro-ph/9706219},
  \adsurl{http://adsabs.harvard.edu/abs/1998MNRAS.296...44G}

\bibitem[\protect\citeauthoryear{Gratton, Lewis \& Efstathiou}{Gratton
  et~al.}{2008}]{Gratton_2007}
Gratton S.,  Lewis A.,    Efstathiou G.,  2008, Phys. Rev., D77, 083507,
  \eprint{0705.3100},
  \adsurl{http://adsabs.harvard.edu/abs/2008PhRvD..77h3507G}

\bibitem[\protect\citeauthoryear{{Hall} \& {Challinor}}{{Hall} \&
  {Challinor}}{2012}]{Hall_2012}
{Hall} A.~C.,  {Challinor} A.,  2012, MNRAS, p.~3504, \eprint{1205.6172},
  \adsurl{http://adsabs.harvard.edu/abs/2012MNRAS.tmp.3504H}

\bibitem[\protect\citeauthoryear{{Hannestad}, {Haugb{\o}lle} \&
  {Schultz}}{{Hannestad} et~al.}{2012}]{Hannestad_2012}
{Hannestad} S.,  {Haugb{\o}lle} T.,    {Schultz} C.,  2012, \jcap, 2, 45,
  \eprint{1110.1257},
  \adsurl{http://adsabs.harvard.edu/abs/2012JCAP...02..045H}

\bibitem[\protect\citeauthoryear{{Ichiki}, {Takada} \& {Takahashi}}{{Ichiki}
  et~al.}{2009}]{Ichiki_2009}
{Ichiki} K.,  {Takada} M.,    {Takahashi} T.,  2009, \prd, 79, 023520,
  \eprint{0810.4921},
  \adsurl{http://adsabs.harvard.edu/abs/2009PhRvD..79b3520I}

\bibitem[\protect\citeauthoryear{{Jimenez}, {Kitching}, {Pe{\~n}a-Garay} \&
  {Verde}}{{Jimenez} et~al.}{2010}]{Jimenez}
{Jimenez} R.,  {Kitching} T.,  {Pe{\~n}a-Garay} C.,    {Verde} L.,  2010,
  \jcap, 5, 35, \eprint{1003.5918},
  \adsurl{http://adsabs.harvard.edu/abs/2010JCAP...05..035J}

\bibitem[\protect\citeauthoryear{{Kaplinghat}, {Knox} \& {Song}}{{Kaplinghat}
  et~al.}{2003}]{Kaplinghat_2003}
{Kaplinghat} M.,  {Knox} L.,    {Song} Y.-S.,  2003, Physical Review Letters,
  91, 241301, \eprint{astro-ph/0303344},
  \adsurl{http://adsabs.harvard.edu/abs/2003PhRvL..91x1301K}

\bibitem[\protect\citeauthoryear{{Komatsu} et~al.,}{{Komatsu}
  et~al.}{2011}]{WMAP7}
{Komatsu} E.,  et~al., 2011, \apjs, 192, 18, \eprint{1001.4538},
  \adsurl{http://adsabs.harvard.edu/abs/2011ApJS..192...18K}

\bibitem[\protect\citeauthoryear{{Kraus}, {Bornschein}, {Bonn}, {Bornschein},
  {Flatt}, {Kovalik}, {M{\"u}ller}, {Otten}, {Schall}, {Th{\"u}mmler} \&
  {Weinheimer}}{{Kraus} et~al.}{2005}]{Kraus_2005}
{Kraus} C.,  {Bornschein} L.,  {Bonn} J.,  {Bornschein} B.,  {Flatt} B.,
  {Kovalik} A.,  {M{\"u}ller} B.,  {Otten} E.~W.,  {Schall} J.~P.,
  {Th{\"u}mmler} T.,    {Weinheimer} C.,  2005, Nuclear Physics B Proceedings
  Supplements, 143, 499,
  \adsurl{http://adsabs.harvard.edu/abs/2005NuPhS.143..499K}

\bibitem[\protect\citeauthoryear{{Lesgourgues}, {Matarrese}, {Pietroni} \&
  {Riotto}}{{Lesgourgues} et~al.}{2009}]{Lesgourgues_2009}
{Lesgourgues} J.,  {Matarrese} S.,  {Pietroni} M.,    {Riotto} A.,  2009,
  \jcap, 6, 17, \eprint{0901.4550},
  \adsurl{http://adsabs.harvard.edu/abs/2009JCAP...06..017L}

\bibitem[\protect\citeauthoryear{{Lesgourgues} \& {Pastor}}{{Lesgourgues} \&
  {Pastor}}{2006}]{Lesgourgues_2006}
{Lesgourgues} J.,  {Pastor} S.,  2006, \physrep, 429, 307,
  \eprint{astro-ph/0603494},
  \adsurl{http://adsabs.harvard.edu/abs/2006PhR...429..307L}

\bibitem[\protect\citeauthoryear{{Lesgourgues} \& {Tram}}{{Lesgourgues} \&
  {Tram}}{2011}]{CLASS_neutrinos}
{Lesgourgues} J.,  {Tram} T.,  2011, \jcap, 9, 32, \eprint{1104.2935},
  \adsurl{http://adsabs.harvard.edu/abs/2011JCAP...09..032L}

\bibitem[\protect\citeauthoryear{{Lewis} \& {Challinor}}{{Lewis} \&
  {Challinor}}{2002}]{CAMB_neutrinos}
{Lewis} A.,  {Challinor} A.,  2002, \prd, 66, 023531,
  \eprint{astro-ph/0203507},
  \adsurl{http://adsabs.harvard.edu/abs/2002PhRvD..66b3531L}

\bibitem[\protect\citeauthoryear{{Ma} \& {Bertschinger}}{{Ma} \&
  {Bertschinger}}{1995}]{Ma_1995}
{Ma} C.-P.,  {Bertschinger} E.,  1995, ApJ, 455, 7, \eprint{astro-ph/9506072},
  \adsurl{http://adsabs.harvard.edu/abs/1995ApJ...455....7M}

\bibitem[\protect\citeauthoryear{McDonald et~al.,}{McDonald
  et~al.}{2006}]{McDonald_2004data}
McDonald P.,  et~al., 2006, ApJS, 163, 80, \eprint{astro-ph/0405013},
  \adsurl{http://adsabs.harvard.edu/abs/2006ApJS..163...80M}

\bibitem[\protect\citeauthoryear{{Mao}, {Tegmark}, {McQuinn}, {Zaldarriaga} \&
  {Zahn}}{{Mao} et~al.}{2008}]{Mao_2008}
{Mao} Y.,  {Tegmark} M.,  {McQuinn} M.,  {Zaldarriaga} M.,    {Zahn} O.,  2008,
  \prd, 78, 023529, \eprint{0802.1710},
  \adsurl{http://adsabs.harvard.edu/abs/2008PhRvD..78b3529M}

\bibitem[\protect\citeauthoryear{{Marulli}, {Carbone}, {Viel}, {Moscardini} \&
  {Cimatti}}{{Marulli} et~al.}{2011}]{Marulli_2011}
{Marulli} F.,  {Carbone} C.,  {Viel} M.,  {Moscardini} L.,    {Cimatti} A.,
  2011, MNRAS, 418, 346, \eprint{1103.0278},
  \adsurl{http://adsabs.harvard.edu/abs/2011MNRAS.418..346M}

\bibitem[\protect\citeauthoryear{{McQuinn}, {Zahn}, {Zaldarriaga}, {Hernquist}
  \& {Furlanetto}}{{McQuinn} et~al.}{2006}]{McQuinn_2006}
{McQuinn} M.,  {Zahn} O.,  {Zaldarriaga} M.,  {Hernquist} L.,    {Furlanetto}
  S.~R.,  2006, ApJ, 653, 815, \eprint{astro-ph/0512263},
  \adsurl{http://adsabs.harvard.edu/abs/2006ApJ...653..815M}

\bibitem[\protect\citeauthoryear{{Planck Collaboration}}{{Planck
  Collaboration}}{2005}]{Planck}
{Planck Collaboration} 2005, ESA-SCI, \eprint{astro-ph/0604069},
  \adsurl{http://adsabs.harvard.edu/abs/2006astro.ph..4069T}

\bibitem[\protect\citeauthoryear{{Ringwald} \& {Wong}}{{Ringwald} \&
  {Wong}}{2004}]{Ringwald_Wong_2004}
{Ringwald} A.,  {Wong} Y.~Y.~Y.,  2004, \jcap, 12, 5, \eprint{hep-ph/0408241},
  \adsurl{http://adsabs.harvard.edu/abs/2004JCAP...12..005R}

\bibitem[\protect\citeauthoryear{{Saito}, {Takada} \& {Taruya}}{{Saito}
  et~al.}{2008}]{Saito_2008}
{Saito} S.,  {Takada} M.,    {Taruya} A.,  2008, Physical Review Letters, 100,
  191301, \eprint{0801.0607},
  \adsurl{http://adsabs.harvard.edu/abs/2008PhRvL.100s1301S}

\bibitem[\protect\citeauthoryear{{Saito}, {Takada} \& {Taruya}}{{Saito}
  et~al.}{2009}]{Saito_2009}
{Saito} S.,  {Takada} M.,    {Taruya} A.,  2009, \prd, 80, 083528,
  \eprint{0907.2922},
  \adsurl{http://adsabs.harvard.edu/abs/2009PhRvD..80h3528S}

\bibitem[\protect\citeauthoryear{{Schlegel}, {White} \&
  {Eisenstein}}{{Schlegel} et~al.}{2009}]{Schlegel_2009}
{Schlegel} D.,  {White} M.,    {Eisenstein} D.,  2009, in Astro2010, A\&A
  Decadal Survey. {The Baryon Oscillation Spectroscopic Survey: Precision
  measurement of the absolute cosmic distance scale}.
National Academies Press, p.~314, \eprint{0902.4680},
  \adsurl{http://adsabs.harvard.edu/abs/2009astro2010S.314S}

\bibitem[\protect\citeauthoryear{{Scoccimarro}}{{Scoccimarro}}{1998}]{Scoccimarro_1998}
{Scoccimarro} R.,  1998, MNRAS, 299, 1097, \eprint{astro-ph/9711187},
  \adsurl{http://adsabs.harvard.edu/abs/1998MNRAS.299.1097S}

\bibitem[\protect\citeauthoryear{{Seljak}}{{Seljak}}{2000}]{Seljak_2000}
{Seljak} U.,  2000, MNRAS, 318, 203, \eprint{astro-ph/0001493},
  \adsurl{http://adsabs.harvard.edu/abs/2000MNRAS.318..203S}

\bibitem[\protect\citeauthoryear{{Seljak}, {Slosar} \& {McDonald}}{{Seljak}
  et~al.}{2006}]{Seljak_2006}
{Seljak} U.,  {Slosar} A.,    {McDonald} P.,  2006, \jcap, 10, 14,
  \eprint{astro-ph/0604335},
  \adsurl{http://adsabs.harvard.edu/abs/2006JCAP...10..014S}

\bibitem[\protect\citeauthoryear{{Shoji} \& {Komatsu}}{{Shoji} \&
  {Komatsu}}{2009}]{Shoji_Komatsu_2009}
{Shoji} M.,  {Komatsu} E.,  2009, ApJ, 700, 705, \eprint{0903.2669},
  \adsurl{http://adsabs.harvard.edu/abs/2009ApJ...700..705S}

\bibitem[\protect\citeauthoryear{{Shoji} \& {Komatsu}}{{Shoji} \&
  {Komatsu}}{2010}]{Shoji_Komatsu_2010}
{Shoji} M.,  {Komatsu} E.,  2010, \prd, 81, 123516,
  \adsurl{http://adsabs.harvard.edu/abs/2010PhRvD..81l3516S}

\bibitem[\protect\citeauthoryear{{Singh} \& {Ma}}{{Singh} \&
  {Ma}}{2003}]{Singh_Ma_2003}
{Singh} S.,  {Ma} C.-P.,  2003, \prd, 67, 023506, \eprint{astro-ph/0208419},
  \adsurl{http://adsabs.harvard.edu/abs/2003PhRvD..67b3506S}

\bibitem[\protect\citeauthoryear{{Springel}}{{Springel}}{2005}]{Springel_2005}
{Springel} V.,  2005, MNRAS, 364, 1105, \eprint{astro-ph/0206393},
  \adsurl{http://adsabs.harvard.edu/abs/2003MNRAS.339..289S}

\bibitem[\protect\citeauthoryear{{Takada}, {Komatsu} \& {Futamase}}{{Takada}
  et~al.}{2006}]{Takada_Komatsu_2006}
{Takada} M.,  {Komatsu} E.,    {Futamase} T.,  2006, \prd, 73, 083520,
  \eprint{astro-ph/0512374},
  \adsurl{http://adsabs.harvard.edu/abs/2006PhRvD..73h3520T}

\bibitem[\protect\citeauthoryear{Vallinotto, Viel, Das \& Spergel}{Vallinotto
  et~al.}{2009}]{Vallinotto_2009}
Vallinotto A.,  Viel M.,  Das S.,    Spergel D.~N.,  2009, ApJ,
  \eprint{0910.4125},
  \adsurl{http://adsabs.harvard.edu/abs/2011ApJ...735...38V}

\bibitem[\protect\citeauthoryear{{Viel}, {Haehnelt} \& {Springel}}{{Viel}
  et~al.}{2010}]{Viel_2010}
{Viel} M.,  {Haehnelt} M.~G.,    {Springel} V.,  2010, \jcap, 6, 15,
  \eprint{1003.2422},
  \adsurl{http://adsabs.harvard.edu/abs/2010JCAP...06..015V}

\bibitem[\protect\citeauthoryear{{Vikhlinin}, {Kravtsov}, {Burenin}, {Ebeling},
  {Forman}, {Hornstrup}, {Jones}, {Murray}, {Nagai}, {Quintana} \&
  {Voevodkin}}{{Vikhlinin} et~al.}{2009}]{Vikhlinin_2009}
{Vikhlinin} A.,  {Kravtsov} A.~V.,  {Burenin} R.~A.,  {Ebeling} H.,  {Forman}
  W.~R.,  {Hornstrup} A.,  {Jones} C.,  {Murray} S.~S.,  {Nagai} D.,
  {Quintana} H.,    {Voevodkin} A.,  2009, ApJ, 692, 1060, \eprint{0812.2720},
  \adsurl{http://adsabs.harvard.edu/abs/2009ApJ...692.1060V}

\bibitem[\protect\citeauthoryear{{Villaescusa-Navarro}, {Miralda-Escud{\'e}},
  {Pe{\~n}a-Garay} \& {Quilis}}{{Villaescusa-Navarro} et~al.}{2011}]{Paco_2011}
{Villaescusa-Navarro} F.,  {Miralda-Escud{\'e}} J.,  {Pe{\~n}a-Garay} C.,
  {Quilis} V.,  2011, \jcap, 6, 27, \eprint{1104.4770},
  \adsurl{http://adsabs.harvard.edu/abs/2011JCAP...06..027V}

\bibitem[\protect\citeauthoryear{{Wagner}, {Verde} \& {Jimenez}}{{Wagner}
  et~al.}{2012}]{Wagner_2012}
{Wagner} C.,  {Verde} L.,    {Jimenez} R.,  2012, ApJ, 752, L31,
  \eprint{1203.5342},
  \adsurl{http://adsabs.harvard.edu/abs/2012ApJ...752L..31W}

\bibitem[\protect\citeauthoryear{Wang, Haiman, Hu, Khoury \& May}{Wang
  et~al.}{2005}]{Wang_2005}
Wang S.,  Haiman Z.,  Hu W.,  Khoury J.,    May M.,  2005, Phys. Rev. Lett.,
  95, 011302, \eprint{astro-ph/0505390},
  \adsurl{http://adsabs.harvard.edu/abs/2005PhRvL..95a1302W}

\bibitem[\protect\citeauthoryear{{Weinberg}}{{Weinberg}}{2008}]{Weinberg_book}
{Weinberg} S.,  2008, \emph{Cosmology}.
Oxford University Press

\bibitem[\protect\citeauthoryear{Weisstein}{Weisstein}{2012a}]{Legendre}
Weisstein E.~W.,  2012a, \emph{Legendre Polynomial}.
From MathWorld--A Wolfram Web Resource

\bibitem[\protect\citeauthoryear{Weisstein}{Weisstein}{2012b}]{Bessel}
Weisstein E.~W.,  2012b, \emph{Spherical Bessel Function of the First Kind}.
From MathWorld--A Wolfram Web Resource

\bibitem[\protect\citeauthoryear{{Wong}}{{Wong}}{2008}]{Wong_2008}
{Wong} Y.~Y.~Y.,  2008, \jcap, 10, 35, \eprint{0809.0693},
  \adsurl{http://adsabs.harvard.edu/abs/2008JCAP...10..035W}

\bibitem[\protect\citeauthoryear{{Wong}}{{Wong}}{2011}]{Wong_review}
{Wong} Y.~Y.~Y.,  2011, Annual Review of Nuclear and Particle Science, 61, 69,
  \eprint{1111.1436},
  \adsurl{http://adsabs.harvard.edu/abs/2011ARNPS..61...69W}

\bibitem[\protect\citeauthoryear{{Xia}, {Granett}, {Viel}, {Bird}, {Guzzo},
  {Haehnelt}, {Coupon}, {McCracken} \& {Mellier}}{{Xia}
  et~al.}{2012}]{Xia_2012}
{Xia} J.-Q.,  {Granett} B.~R.,  {Viel} M.,  {Bird} S.,  {Guzzo} L.,  {Haehnelt}
  M.~G.,  {Coupon} J.,  {McCracken} H.~J.,    {Mellier} Y.,  2012, \jcap, 6,
  10, \eprint{1203.5105},
  \adsurl{http://adsabs.harvard.edu/abs/2012JCAP...06..010X}

\end{thebibliography}

\end{document}